\documentclass[twocolumn,showpacs,preprintnumbers,amsmath,amssymb,superscriptaddress]{revtex4}

\usepackage{graphicx}
\usepackage{dcolumn}
\usepackage{bm}

\def\bea{\begin{eqnarray}}
\def\eea{\end{eqnarray}}

\def\pp{\mbox{$p$-$p$} }
\def\auau{\mbox{Au-Au} }

\def\aa{\mbox{A-A} }
\def\nn{\mbox{N-N} }

\def\deta{$\eta_\Delta$ }

\begin{document} 


\preprint{Version 1.4}

\title{
On the presence  of nonjet ``higher harmonic'' components in \\ 2D angular correlations from high energy heavy ion collisions
}

\author{Thomas A.\ Trainor}\affiliation{CENPA 354290, University of Washington, Seattle, WA 98195}
\author{Duncan J.\ Prindle}\affiliation{CENPA 354290, University of Washington, Seattle, WA 98195}
\author{R.\ L.\ Ray}\affiliation{Department of Physics, University of Texas at Austin, Austin, TX 78712}


\date{\today}

\begin{abstract}
{\bf Background:} 
Possible evidence for triangular flow and other ``higher harmonic flows'' in high energy nucleus-nucleus collisions has received much attention recently. 
It is conjectured that several higher harmonic flows $v_m$ may result from initial-state geometry fluctuations in \aa collisions coupled to a radially-expanding medium. But as with ``elliptic flow'' $v_2$ measurements, non-hydrodynamic mechanisms such as jet production may contribute to  other higher azimuth multipoles $v_m$ as biases.
{\bf Purpose:}
Careful distinctions should be maintained between jet-related and nonjet (possibly hydrodynamic) contributions to $v_m$ (e.g., ``nonflow'' and ``flow''). In this study we consider several questions: (a) To what extent do jet-like structures in two-dimensional (2D) angular correlations contribute to azimuth multipoles inferred from various $v_m$ methods? (b) If a multipole element is added to a 2D fit model is a nonzero amplitude indicative of a corresponding flow component? and (c) Can 2D correlations establish the necessity of nonjet contributions to some or all higher multipoles?
{\bf Method:} 
Model fits to 2D angular correlations are used to establish the origins of azimuth multipoles inferred from 1D projections onto azimuth or from nongraphical numerical methods.
{\bf Results:} We find that jet-like angular correlations, and specifically a 2D peak at the angular origin consistent with jet production, constitute the dominant contribution to inferred higher multipoles, and the data do not {\em require} higher multipoles in isolation from the jet-like 2D peak.
{\bf Conclusions:}
Inference of ``higher harmonic flows'' results from identifying certain nominally jet-like structure as flow manifestations through unjustified application of 1D Fourier series analysis. Although the peak structure at the angular origin is strongly modified in more-central collisions some properties remain compatible with relevant pQCD theory expectations for jet production.
\end{abstract}


\maketitle

 \section{Introduction}

Nuclear collisions at the Relativistic Heavy Ion Collider (RHIC)  have been interpreted within a  hydrodynamic (hydro) context in terms of formation of a thermalized, flowing partonic medium with small viscosity~\cite{qgp1,qgp2}. 
But alternative analysis of spectrum and correlation data has revealed a minimum-bias jet contribution whose variation with \aa centrality and collision energy seems to conflict with hydro expectations~\cite{axialci,anomalous,hardspec,fragevo,jetsyield}.  
In particular, a large-amplitude same-side (SS) 2D peak centered at the origin in angular correlations on relative pseudorapidity $\eta$ and azimuth $\phi$, expected as a component of jet manifestations, persists even in central \auau collisions, albeit the SS peak is elongated on $\eta$ relative to a nominally symmetric jet cone~\cite{axialci,anomalous}. 

In response the jet interpretation of the SS 2D peak has been challenged in the literature. One strategy was based on introduction of ``triangular flow'' to angular correlations. According to a conjecture in Ref.~\cite{gunther} triangular flow should correspond to ``triangularity'' of the event-wise initial-state \aa overlap geometry. Triangular flow would explain both the SS peak and AS double peak in more-central \auau collisions~\cite{asdouble} previously attributed to Mach cones~\cite{mach}. In a more-recent strategy ``higher harmonic flows'' were invoked to explain all aspects of the $\eta$-elongated SS peak or ``soft ridge''~\cite{luzum}. The various harmonic flows were attributed to fluctuations in the initial-state \aa geometry coupled to radial expansion of the bulk medium~\cite{sorensen,flowflucts}. In an alternative scheme the elongated SS 2D peak was explained in terms of  glasma flux tubes coupled to radial flow~\cite{glasma,tglasma}.

The problem of cylindrical multipoles (Fourier series coefficients) derived from 1D projection of all 2D angular correlations onto 1D azimuth was considered in Ref.~\cite{multipoles}.  That study concluded that in almost all cases claimed higher harmonic flows (Fourier series index $m > 2$) are actually 1D Fourier components of the SS 2D peak. The SS peak can be represented accurately by a factorized 2D model function~\cite{anomalous} but {\em not} by a 1D Fourier series due to substantial $\eta$ variation of the peak structure, and some peak properties remain consistent with jet formation. 

In the present study we confront a more subtle issue: If a sextupole Fourier element is added to the ``standard'' 2D fit model described in Ref.~\cite{anomalous} and a nonzero sextupole amplitude results, what does that imply physically? Does a nonzero sextupole amplitude {\em necessarily} imply that ``triangular flow'' plays a role in nuclear collisions? If added multipole model elements result in significant changes to other fit parameters are the data systematics reported in Ref.~\cite{anomalous} arbitrary and misleading?

With recently-reported 2D angular correlation data we demonstrate that attempts to add extra Fourier terms (multipole elements) to the 2D data model of Ref.~\cite{anomalous} are equivalent to minor alteration of the SS 2D peak model, and such modifications are not systematically significant. Substantial variations in several parameter values do arise when the number of model parameters exceeds that required by the correlation structure in the data. However, the parameter changes are strongly correlated and are found to be equivalent to adding a narrow 1D Gaussian to the SS 2D peak model. That addition does not change the {\em combined} data model significantly. 

This paper is arranged as follows: In Sec.~\ref{basic} we introduce some basics of 2D angular correlation measurement. In Secs.~\ref{stdfit} and \ref{modelfit} we define the standard 2D fit model for this study and provide some example fits to data. In Secs.~\ref{1dproj} and \ref{imposing} we consider 1D projections of data onto pseudorapidity $\eta$ and azimuth $\phi$ differences and some results of 1D Fourier analysis on azimuth. In Sec.~\ref{2dsex} we consider consequences of adding a sextupole to the standard 2D fit model within a limited detector $\eta$ acceptance and possible interpretations. In Sec.~\ref{higherharm} we determine what azimuth multipole elements are required by 2D data and what are not required and may be omitted.
Secs.~\ref{disc} and \ref{sum} present the discussion and summary.

 \section{Basic 2D angular correlations} \label{basic}

This study addresses the problem of modeling two-particle angular correlations and physical interpretation of the model elements. Different modeling methods may suggest different physical interpretations. Some imposed terminology presumes specific interpretations. One can question whether there is a unique ``best'' model and whether physical interpretations must depend on analysis techniques or may be uniquely determined by data.

\subsection{Kinematic variables}


Two-particle correlations are structures in pair-density distributions on six-dimensional momentum space $(p_{t1},\eta_1,\phi_1,p_{t2},\eta_2,\phi_2)$. 
2D correlations on transverse momentum $p_t$ or transverse rapidity $y_t = \ln[(p_t + m_t) / m_\pi]$ ($m_\pi$ is assumed for unidentified hadrons) are complementary to 4D angular correlations in the 6D pair momentum space. 
$y_t$ is preferred for visualizing correlation structure on transverse momentum. 
Angular correlations can be measured by integrating over the entire $(p_t,p_t)$ pair acceptance (minimum-bias angular correlations) or over subregions (e.g., ``trigger-associated'' dihadron correlations)~\cite{porter2,porter3}. 

Two-particle {\em angular} correlations are defined on 4D momentum subspace $(\eta_1,\eta_2,\phi_1,\phi_2)$. Within acceptance intervals where correlation structure is approximately invariant on mean azimuth or polar angle (e.g.\ $\eta_\Sigma = \eta_1 + \eta_2$) angular correlations can be {\em projected by averaging} onto difference variables (e.g.\ $\eta_\Delta = \eta_1 - \eta_2$) without loss of information to form {\em angular autocorrelations}~\cite{axialcd,inverse}. 2D subspace ($\eta_\Delta,\phi_\Delta$) is then visualized. 
The pair angular acceptance on azimuth can be separated into a same-side (SS) region ($|\phi_\Delta| < \pi / 2$) and an away-side (AS) region ($|\phi_\Delta| > \pi / 2$). The SS region includes {\em intra}\,jet correlations (hadron pairs within single jets), while the AS region includes {\em inter}\,jet correlations (hadron pairs from back-to-back jet pairs).

Centrality variation of the \aa collision geometry is represented by the Glauber model~\cite{powerlaw}. The model parameters are the number of binary \nn collisions $N_{bin}$ and the number of participant-nucleon pairs $N_{part}/2$ as functions of the fractional \aa cross section $\sigma/\sigma_0$. We also define a {\em mean participant path length} $\nu = 2N_{bin} / N_{part}$. The \aa impact parameter is represented by symbol $b$.

\subsection{Correlation measures}


Correlations can be measured with {\em per-particle} statistic  $\Delta \rho / \sqrt{\rho_{ref}} = \rho_0\, (\langle r \rangle-1)$, where $\Delta \rho= \rho_{sib} - \rho_{ref}$ is the correlated-pair density, $\rho_{sib}$ is the sibling (same-event) pair density, $\rho_{ref}$ is the reference or mixed-event pair density, $\langle  r \rangle$ is the mean sibling/mixed pair-number ratio, and prefactor  $\rho_0 = \bar n_{ch} / \Delta \eta\, \Delta \phi  \approx d^2n_{ch} / d\eta d\phi$ is the charged-particle 2D angular density averaged over some angular acceptance $(\Delta \eta,\Delta \phi)$~\cite{axialci,axialcd}.  
Factorization $\rho_{ref} \approx \rho_0^2$ is assumed, and symbol $\Delta x$ denotes a detector acceptance on  $x$.

The {\em per-particle} quadrupole component of 2D angular correlations resulting from a given $v_2$ analysis method is  $A_Q\{\text{method}\} \equiv \rho_0(b) v_2^2\{method\}(b)$. The quadrupole component can also be measured with  {\em per-pair} statistic $\Delta \rho / \rho_{ref} \rightarrow v_2^2\{method\}$, including total azimuth quadrupole  $v_2^2\{2\}$ (method \{2\} denotes the $m = 2$ quadrupole cosine component of {\em all} 1D azimuth projections). Higher multipoles are represented by $v_m^2\{2\}$. Variation of per-pair correlation measures with \aa centrality is typically dominated by a trivial $1/n_{ch}$ factor, or in the case of $v_m$ a $1/\sqrt{n_{ch}}$ factor.

\subsection{Example 2D data histograms}

\begin{figure}[h]
 \includegraphics[width=1.65in,height=1.5in]{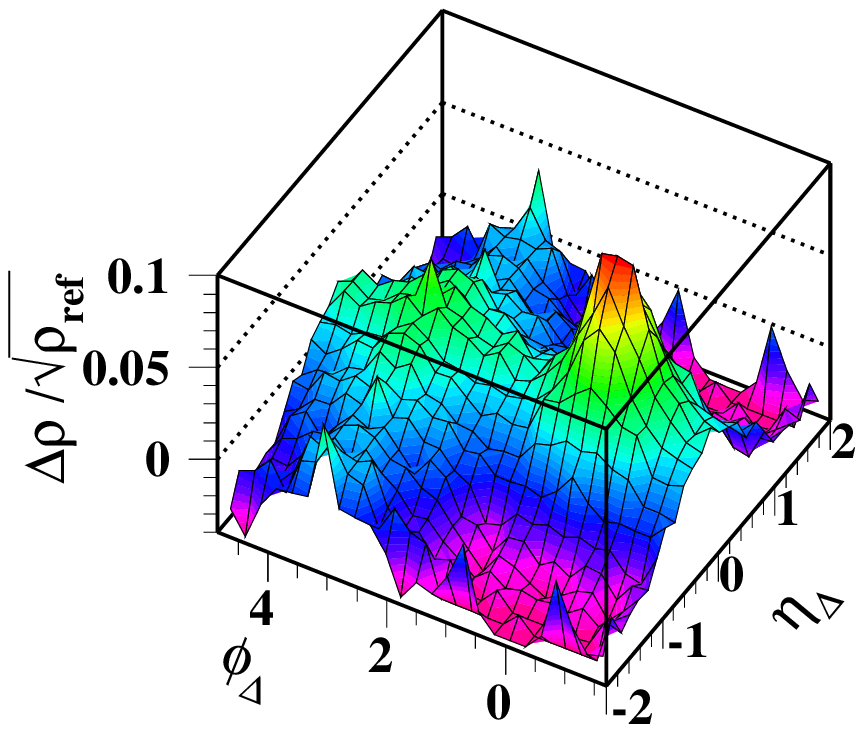}
 \includegraphics[width=1.65in,height=1.5in]{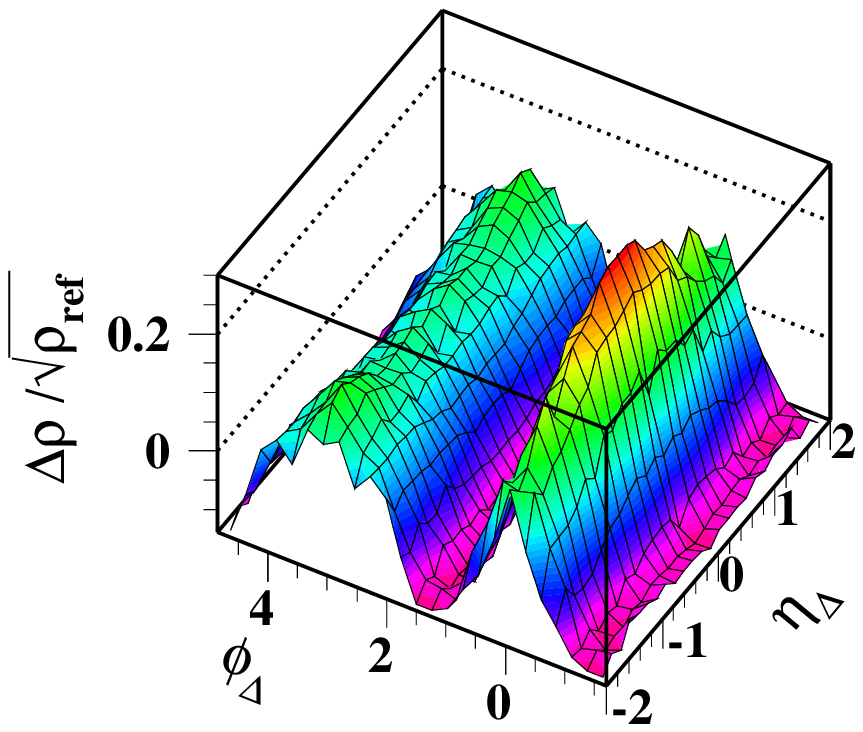}
\caption{\label{basicdata}
$p_t$-integral 2D angular correlations from 200 GeV \auau collisions.
Left: 84-93\%-central collisions (approximately \nn collisions).
Right: 0-5\%-central collisions.
} 
 \end{figure}

Figure~\ref{basicdata} shows examples from analysis of 200 GeV \auau collisions~\cite{anomalous}. On the left are most-peripheral 84-93\% data that approximate \nn (p-p) collisions. On the right are most-central 0-5\% data. These are minimum-bias ($p_t$-integral) data: the only momentum cut is $p_t > 0.15$ GeV/c. There are no ``trigger-associated'' $p_t$ cuts applied. The basic correlation histograms have not been further processed (no ``background'' is subtracted).

Interpretation of the \nn data structure in the left panel is straightforward. There is a broader same-side (SS) 2D peak at the angular origin attributed to minimum-bias jets or minijets. Superposed on that is a narrower 2D peak attributed to Bose-Einstein correlations (BEC) plus conversion-electron pairs. An away-side 1D peak on azimuth (AS ridge) is attributed to back-to-back jet pairs. A 1D peak on $\eta_\Delta$ is attributed to dissociation (fragmentation) of projectile nucleons. No quadrupole $\cos(2\phi_\Delta)$ component is evident in the peripheral data.

Figure~\ref{basicdata} (right panel) includes only two of those features: the SS 2D peak and the AS 1D peak. For those most-central \auau data the azimuth quadrupole has a negligible amplitude. The SS peak dominates angular correlations in all cases. Whereas the SS peak in \nn collisions is elongated 2:1 in the azimuth direction the SS peak in central \auau collisions becomes elongated 3:1 in the pseudorapidity direction. The $\eta$-elongated SS peak has been referred to as a ``soft ridge.'' Although there is no {\em formal} difference between the peak shapes for two limiting centralities (both are 2D Gaussians) there is a strong preference in the heavy-ion literature to interpret the elongated SS peak in more-central collisions in terms of flows.
In this analysis we focus on centrality evolution and model descriptions of the SS 2D peak, and its manifestations in various ``higher harmonic flow'' measurements.

\subsection{Correlation structure terminology} \label{terminology}


Measured angular correlations on $(\eta_\Delta,\phi_\Delta)$ include only a few components and can be modeled accurately by simple functional forms. In this study we distinguish between {\em components} of measured angular correlations and {\em elements} of 2D fit models. Two of the fit-model elements (SS 2D peak, AS 1D peak) have been interpreted in terms of measured and theoretical jet systematics~\cite{axialci,anomalous,jetsyield}. One element (quadrupole) has been interpreted in terms of ``elliptic flow,'' a conjectured hydro phenomenon. 

Conventional references to correlation structure may be confusing. In an assumed hydro context correlation structures are categorized as either flows or nonflows. But a jet interpretation for some structures is compelling. Jets are expected to appear in descriptions of high-energy nuclear collisions, at least in \pp and more-peripheral \aa collisions. In contrast, the systematics of nominally flow-related structures appear to contradict some aspects of hydro theory~\cite{davidhq,davidhq2,quadspec}. We therefore elect to categorize correlation structure as either jet-related or nonjet. We continue to use the same terminology for all \auau centralities, although the jet {\em interpretation} for some structures may be questioned in more-central collisions.

Jet-related structure then denotes SS 2D and AS 1D peaks. The former is modified in more-central \auau collisions, but the question remains whether a flow description becomes appropriate there. Nonjet structure includes the BEC-electron peak, the 1D Gaussian on $\eta_\Delta$ (projectile dissociation) and a quadrupole unrelated to SS 2D and AS 1D peaks. Nonjet quadrupole systematics clearly demonstrate  strong correlation with the initial-state \aa geometry. Evidence in data for other nonjet multipoles is the principal subject of the present study.

\subsection{Fit-model criteria for 2D histograms}

Attempts to isolate 2D or 1D correlation structures from a (large) combinatoric background have generally followed one of two methods: 
(a) ZYAM (zero yield at minimum) subtraction from 1D dihadron correlations on azimuth~\cite{asdouble} and (b) model fits to 2D histograms~\cite{axialci,anomalous}. 
Method (a) depends on background estimation, and in particular measurements of parameter $v_2$ by nongraphical numerical methods which may include substantial jet-related bias~\cite{davidhq,gluequad,davidhq2}. If the subtracted background is consequently  biased the inferred jet-related structure may be underestimated and distorted~\cite{tzyam}.

Method (b) requires  definition of a 2D fit model function including several elements. A range of possibilities exists, some choices being motivated by physical models or expectations for observation of certain phenomena. We prefer a mathematical model that is not motivated by physical models or expectations but does satisfy a few general requirements for statistical models:
The same model should apply to all collision centralities, not just restricted intervals. The number of model parameters should not exceed what is {\em required} by the structure in the data (parsimony). 
Each model element should be established as {\em necessary}\,: omission of that element should result in a substantially worsened fit with equivalent residuals structure. A {\em sufficient} model is a sum of necessary elements that results in negligible residuals structure. A {\em necessary and sufficient} model is ideal.

 \section{Standard 2D fit model} \label{stdfit}

All minimum-bias ($p_t$-integral) 2D angular correlations from \auau collisions at $\sqrt{s_{NN}} =$ 62 and 200 GeV are observed to include three main components: (a) a same-side (SS) 2D peak at the origin on $(\eta_\Delta,\phi_\Delta)$ well approximated by a 2D Gaussian, (b) an away-side (AS) 1D peak (or ``ridge'') on azimuth well approximated by AS azimuth dipole $ \cos(\phi_\Delta)$ and uniform on $\eta_\Delta$ within a few percent (having negligible curvature) over the angular acceptance of the STAR TPC ($\Delta \eta = 2$, $\Delta \phi = 2\pi$), and (c) nonjet (NJ) azimuth quadrupole $\cos(2\phi_\Delta)$ also uniform on \deta to a few percent. 
Model elements (a) and (b) have been interpreted together as representing minimum-bias jets or ``minijets''~\cite{fragevo,jetsyield, anomalous}. Element (c) has been conventionally attributed to elliptic flow, a hydrodynamic phenomenon~\cite{2004}. However, alternative interpretations may be possible~\cite{gluequad}.

Given that data structure the ``standard'' 2D model function for more-central \auau collisions is~\cite{axialci,anomalous,davidhq}
\bea \label{estructfit}
\frac{\Delta \rho}{\sqrt{\rho_{ref}}} \hspace{-.02in}  & = &  \hspace{-.02in}
A_0+  A_{1} \, \exp \left\{- \frac{1}{2} \left[ \left( \frac{\phi_{\Delta}}{ \sigma_{\phi_{\Delta}}} \right)^2 \hspace{-.05in}  + \left( \frac{\eta_{\Delta}}{ \sigma_{\eta_{\Delta}}} \right)^2 \right] \right\} \nonumber \\
&+&  \hspace{-.02in} A_{D}\, \cos(\phi_\Delta - \pi)\}  + \hspace{-.02in}  A_{Q}\, \cos(2\, \phi_\Delta).
\eea
A 1D Gaussian on $\eta_\Delta$ modeling projectile nucleon fragmentation and a 2D exponential modeling Bose-Einstein correlations (BEC) and conversion electron pairs are omitted for simplicity in discussion of more-central \auau collisions where the former has negligible amplitude and the latter is very narrow, restricted to three histogram bins near the origin. The model fits presented here also include a small-amplitude (few percent) modulation of the AS dipole on $\eta_\Delta$ [e.g., panel (a) of Fig.~\ref{2ddatafits}]. Details are omitted here to simplify the presentation. The modulation is described in Sec.\ VII C of Ref.~\cite{anomalous}.

The definition of quadrupole measure $A_Q$ is statistically compatible with jet-related measures $A_{1}$ and $A_D$ (all are per-particle measures), permitting quantitative comparisons between jet and nonjet quadrupole systematics (e.g., comparing jet-related and nonjet contributions to $A_Q$). The standard fit model accurately describes most \pp and \auau 2D data (with respect to centralities and $p_t$ cuts). In specific cases (e.g., ultracentral \auau collisions and ``trigger-associated'' high-$p_t$ cuts) certain modifications may be required to the nominally jet-related SS and AS peak models.

For the present study the adopted model-parameter definitions in Eq.~(\ref{estructfit}) are compatible with those in Ref.~\cite{anomalous}. In other, related papers (e.g., \cite{tzyam,multipoles}) certain parameter definitions (e.g., $A_Q$ and $A_D$) differ by factors 2 to be consistent with Fourier series definitions.

 \section{model fits to \auau data} \label{modelfit}

As an example of model fits to 2D angular correlations we present data from 0-5\% central \auau collisions reported in Ref.~\cite{anomalous}. Similar fit quality is obtained for $p_t$-integral correlations from all \auau collision centralities.

\subsection{Typical fit results, 0-5\% central Au-Au}

\begin{figure}[h]
 \includegraphics[width=1.65in]{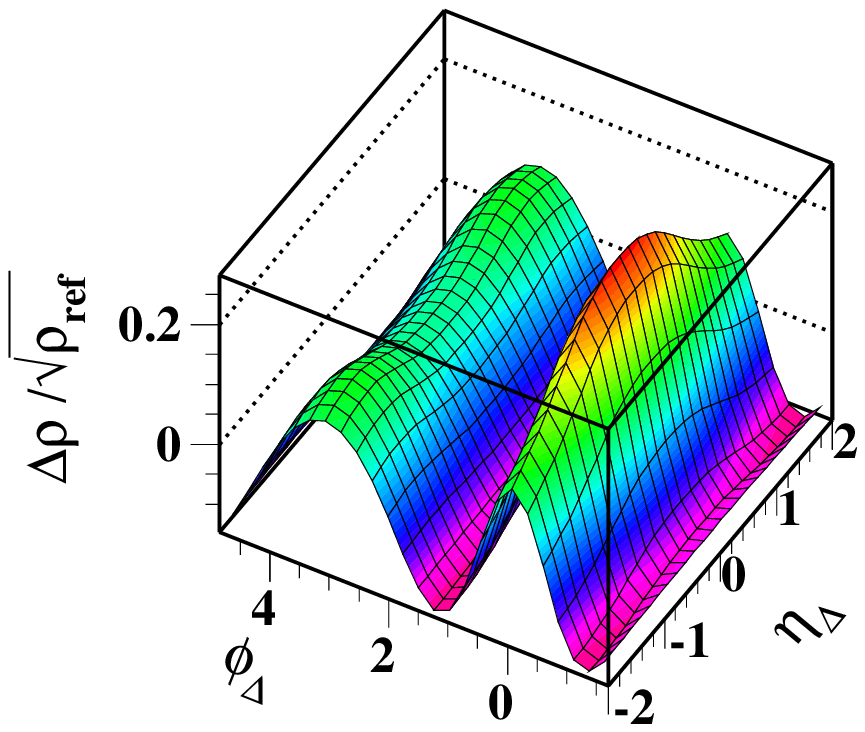}
 \put(-100,80){(a)}
\includegraphics[width=1.65in]{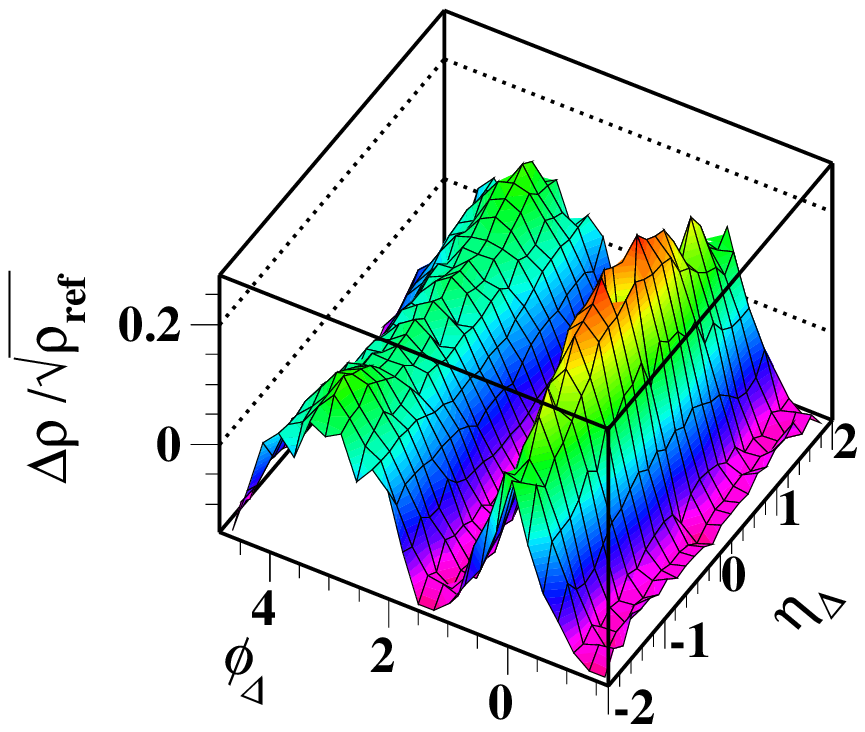}
 \put(-100,80){(b)} \\
 \includegraphics[width=1.65in]{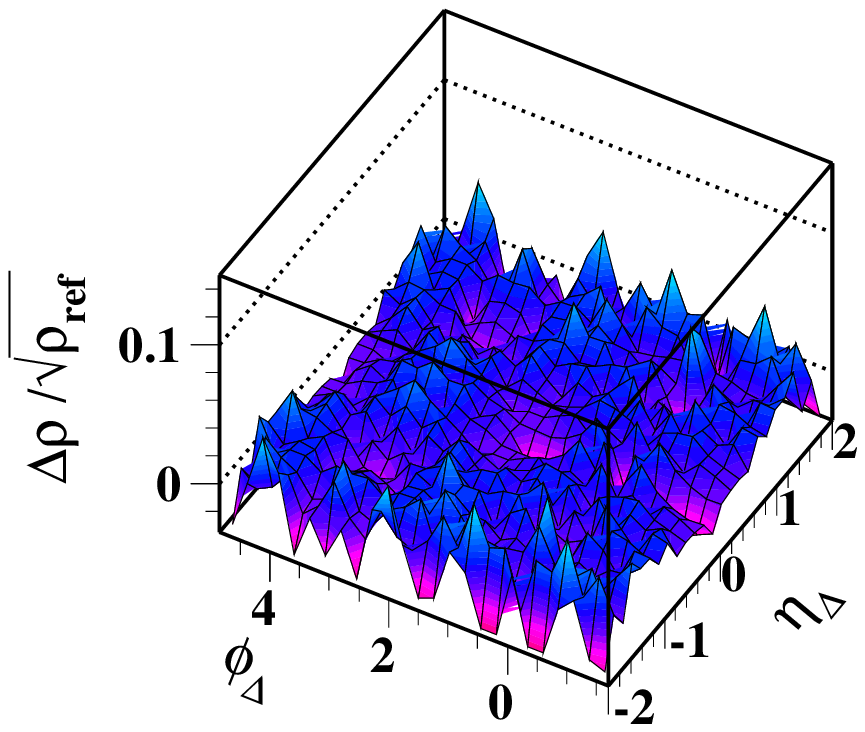}
 \put(-100,80){(c)}
 \includegraphics[width=1.65in]{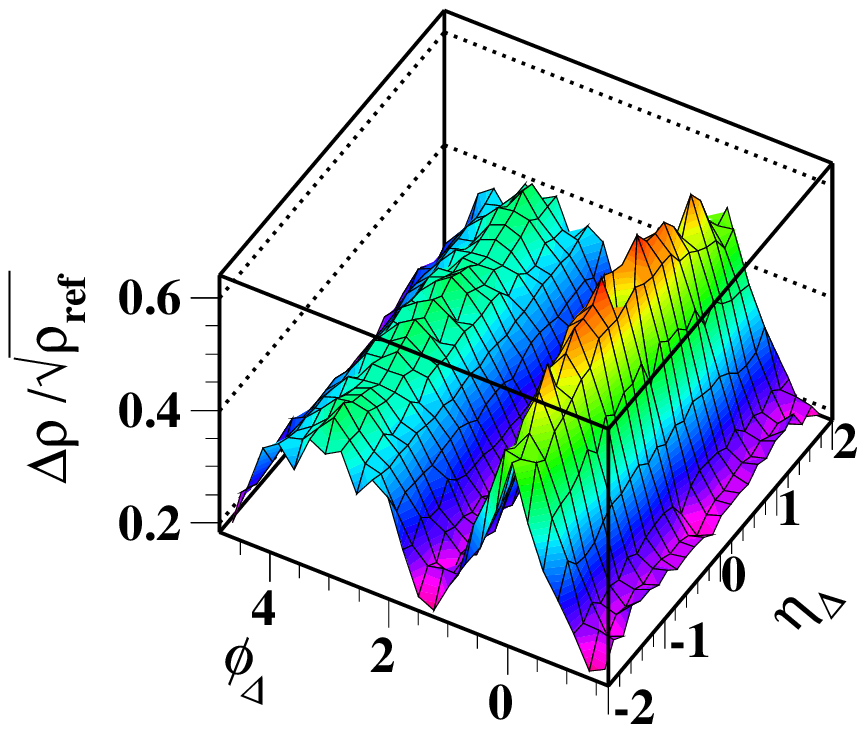}
 \put(-100,80){(d)}
\caption{\label{2ddatafits}
Model fit to 0-5\%-central 200 GeV \auau collisions:
(a) model, (b) data, (c) fit residuals, (d) inferred jet structure. 
} 
 \end{figure}

Figure~\ref{2ddatafits} illustrates typical 2D model fits to angular correlations. The best-fit model (a) is compared to the fitted data (b). In panel (c) are the fit residuals ($2\times$ more-sensitive scale), generally consistent with statistical uncertainties. The BEC-electron peak has been suppressed in these data because it is confined to only three bins at the origin. In more-peripheral collisions the BEC peak must be modeled explicitly.

Panel (d) estimates the true jet contribution (SS 2D and AS 1D peaks). For 0-5\% collisions the nonjet quadrupole amplitude is consistent with zero~\cite{davidhq}. In more-central \auau collisions the AS dipole associated with the AS 1D peak is observed to have a small modulation on $\eta_\Delta$, as discussed in Sec. VII-C of Ref.~\cite{anomalous}. That modulation is included in the 2D fit model for this study and, with the fitted offset, has been subtracted from the data histogram in panel (b) to obtain panel (d).

\subsection{Evolution with \auau centrality and $p_t$}

\begin{figure}[h]
 \includegraphics[width=1.65in,height=1.6in]{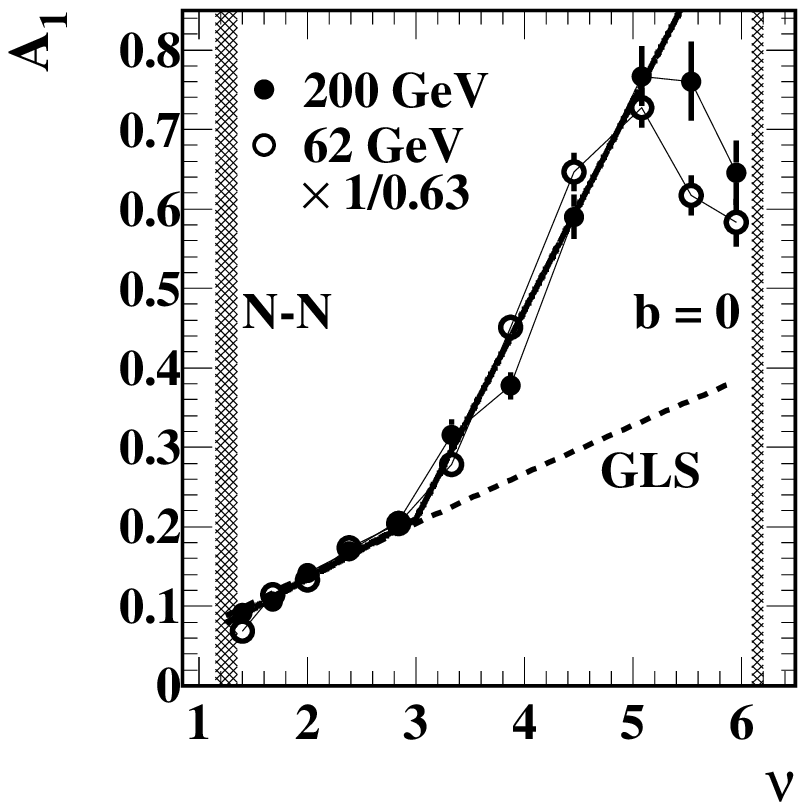}
 \includegraphics[width=1.65in,height=1.6in]{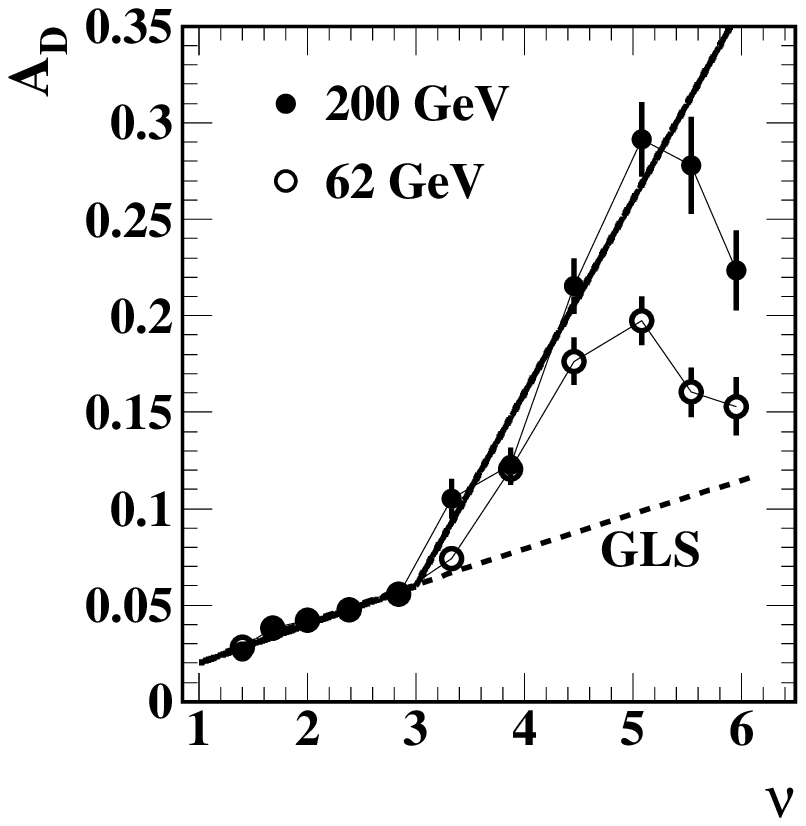}
\caption{\label{centdep}
Variation with \auau centrality for two model parameters. 
Left: Same-side 2D peak amplitude.
Right: Away-side 1D peak amplitude.
 Glauber linear superposition (GLS) trends (dashed curves) correspond to jet production according to \nn binary-collision scaling. A sharp transition (ST) in these jet-related parameter trends near $\nu = 3$ is notable.
} 
 \end{figure}

The full centrality dependence of the model elements is essential for proper physical interpretation of 2D angular correlations. Figure~\ref{centdep} shows the centrality dependence of the SS 2D peak amplitude (left panel) and AS 1D dipole amplitude (right panel)~\cite{anomalous}. The dashed curves represent Glauber linear superposition (GLS) trends expected for jet production in case \aa collisions are linear superpositions of \nn collisions (i.e., transparent). The form of the dashed curves is determined by the Glauber model and amplitudes are matched to the peripheral \auau data. 

These nominally ``jet-related'' \auau structures are observed to follow the GLS reference accurately from \pp (N-N) up to a specific \auau centrality $\nu \approx 3$ where the amplitude trends undergo a {\em sharp transition} (ST). The ST centrality corresponds to approximately 50\% of the total cross section. Those trends and other evidence suggest that jet production is the correct interpretation of the SS 2D and AS 1D correlation structures~\cite{fragevo,jetsyield}. It is notable that the SS 2D peak amplitude scales with energy as $\log(\sqrt{s_{NN}})$ as expected for a QCD process (factor 0.63 in this case) whereas the AS 1D amplitude is unchanging with energy, as expected for dijet production~\cite{anomalous}.


If {\em marginal} $p_t$ cuts are applied ($p_t$ of one particle is constrained but not the other) the shape of the SS 2D peak remains a 2D Gaussian below $p_t \approx 4$ GeV/c, in contrast to so-called trigger-associated $p_t$ cuts for which both particles of a pair are constrained. In the latter case a large fraction of any jet may be excluded from the correlation structure~\cite{fragevo}. Generally speaking, evolution of both SS 2D and AS 1D peak structures with $p_t$ cuts is consistent with jet expectations~\cite{porter2,porter3}. Deviations from the ``standard'' 2D fit model for extreme cases such as ultra-central (0-1\%) collisions are discussed in Sec.~\ref{ultra}.

 \section{Projections onto 1D subspaces} \label{1dproj}

Given the simple structure of 2D angular correlations we focus on the properties of the SS 2D peak and its contribution to various applications of 1D Fourier series. 

\subsection{Isolating the SS 2D (jet-related) peak}

We have established that all 2D histograms for more-central \auau collisions (except ``ultracentral'' 0-1\% data) can be expressed as the sum of an AS dipole ($m=1$), nonjet quadrupole ($m=2$) and SS 2D peak.  The two multipoles are {\em necessary} model elements, as we demonstrate below.
Thus, because of orthogonality of Fourier terms, any ``higher harmonics'' must be derived from the SS 2D peak structure. The  SS 2D peak model may be questioned, but the {\em data} peak can be uniquely isolated with negligible model dependence simply by subtracting the fitted AS dipole and NJ quadrupole terms.

 \begin{figure}[h]
\includegraphics[width=.23\textwidth,height=.21\textwidth]{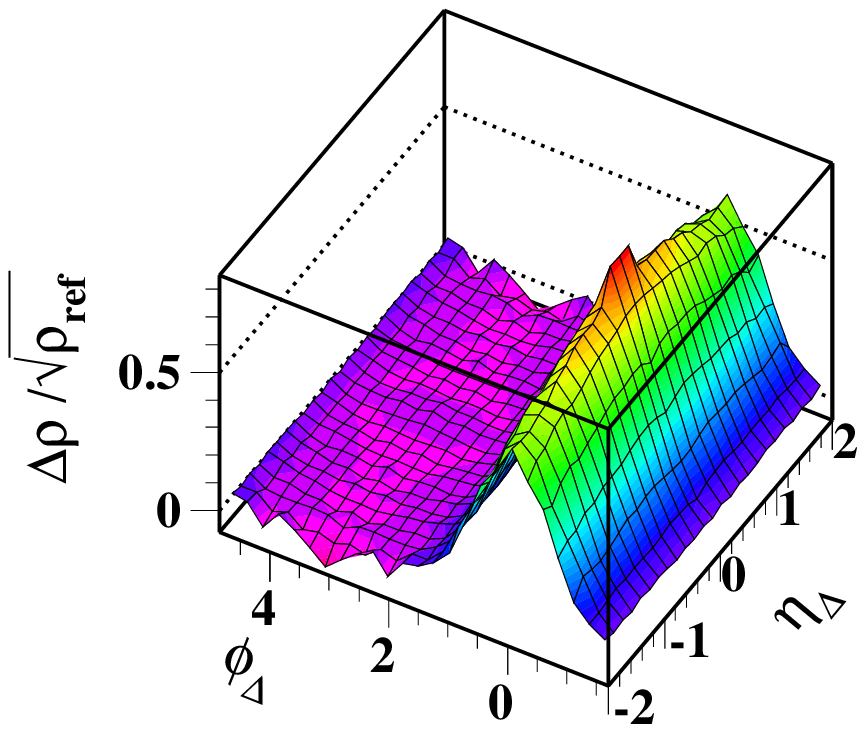}  
\includegraphics[width=.23\textwidth,height=.21\textwidth]{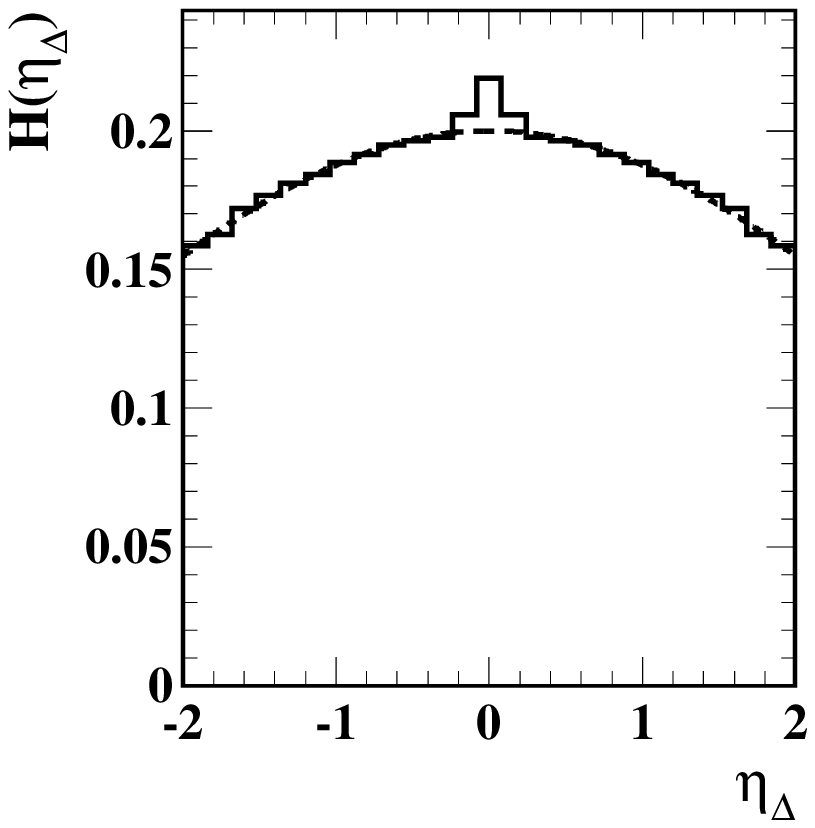}  
\caption{\label{isolate}
Left: Data from a $p_t$-integral analysis of 9-18\%-central 200 GeV \auau collisions from which fitted AS dipole, nonjet quadrupole and  constant offset components have been subtracted, leaving the SS 2D peak.
Right: Projection of the 2D histogram in the left panel onto $\eta_\Delta$. The dashed curve is a Gaussian with amplitude 0.20 and r.m.s.\ width 2.7. 
} 
\ \end{figure}

Figure~\ref{isolate} (left panel) shows the result of subtracting fitted AS 1D dipole and constant offset terms from the 0-5\% central \auau data in Fig.~\ref{2ddatafits} (the narrow BEC-electron peak is truncated in the plot). No statistically-significant AS structure remains, no evidence for ``higher harmonics'' unrelated to the SS 2D peak. 
Similar results are obtained for other centralities. We conclude that the only source of Fourier components with $m > 2$  is the SS 2D peak. 

\subsection{Factorizing the SS 2D peak} \label{factorize}

Because in almost all cases the SS 2D peak is the only significant source of higher multipoles we now consider the SS peak structure. We observe that for all cases the peak is {\em factorizable}. The SS peak can be described accurately by the product $G(\phi_\Delta) H(\eta_\Delta)$. Azimuth factor $G(\phi_\Delta)$ is always  described by a narrow Gaussian. Factor $H(\eta_\Delta)$ is also well approximated in most cases by a Gaussian whose width on $\eta_\Delta$ increases significantly with \auau centrality. In some cases (trigger-associated $p_t$ cuts, higher collision energies) $H(\eta_\Delta)$  may deviate substantially from a Gaussian function. 
To further pursue the higher-harmonics issue we construct projections of the SS 2D peak onto 1D $\eta_\Delta$ and $\phi_\Delta$ for $p_t$-integral data.

Figure~\ref{isolate} (right panel) shows a full $2\pi$ projection of the SS 2D peak in the left panel onto $\eta_\Delta$. The small and narrow peak is BEC plus conversion electrons. The Gaussian (dashed) curve passing through the data histogram is $0.20\exp[-(\eta_\Delta/2.7)^2 /2 ]$.  The SS peak for $p_t$-integral data is  well described by a single 1D Gaussian on $\eta_\Delta$. The same is true for most {\em marginal} $p_t$-differential data.

 \begin{figure}[h]
 \includegraphics[width=1.65in,height=1.6in]{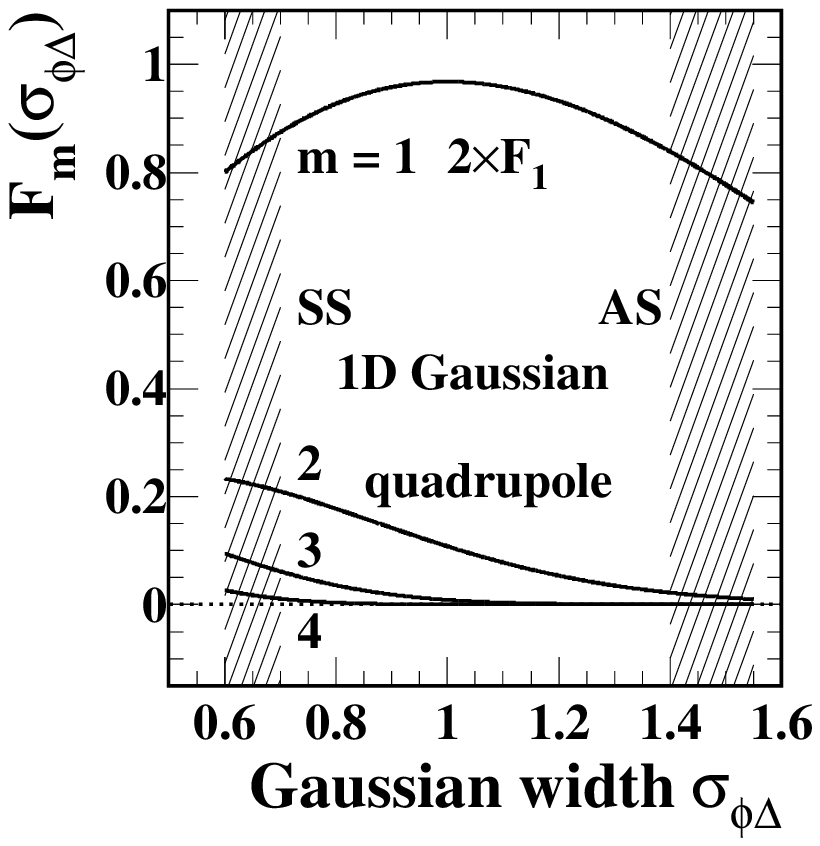}
 \includegraphics[width=1.65in,height=1.6in]{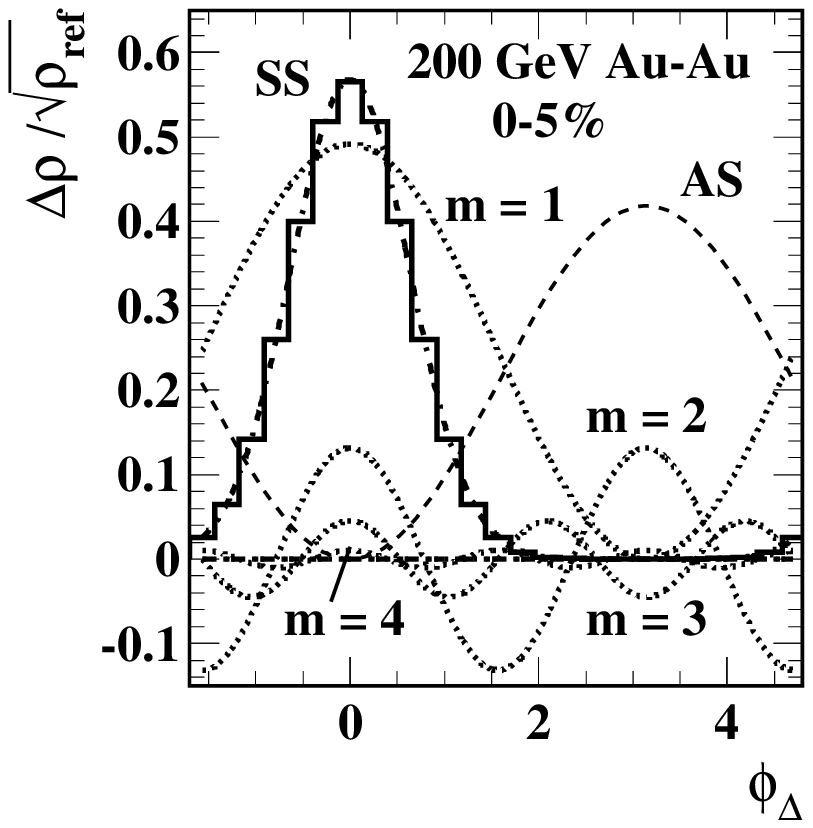}
\caption{\label{fourierfig}
Left: Fourier coefficients for a periodic array of unit-amplitude 1D Gaussians on azimuth as a function of the r.m.s.\ width. The hatched bands correspond to typical peak widths for same-side (SS) and away-side (AS) jet-related peaks.
Right: The histogram is a projection of the SS peak in Fig.~\ref{isolate} (left panel) onto azimuth. The dash-dotted and dashed curves are the SS 1D Gaussian and AS dipole from the 2D model fit. The bold dotted curves are the  SS multipoles inferred from Eq.~(\ref{fourier}).
} 
 \end{figure}

Figure~\ref{fourierfig} (left panel) shows the Fourier coefficients of a periodic array of unit-amplitude 1D Gaussians as a function of its width~\cite{tzyam}. The coefficients are given by
\bea \label{fourier}
F_m(\sigma_{\phi_\Delta}) = \sqrt{2/\pi}\, \sigma_{\phi_\Delta} \exp\{-m^2 \sigma_{\phi_\Delta}^2 / 2\}.
\eea
There are two relevant width regimes, for SS 2D and AS 1D jet-related peaks, denoted by the hatched bands. In the former case several multipoles are relevant. In the latter case, especially for $p_t$-integral data, only the $m=1$ AS dipole is relevant. If ``high-$p_t$'' cuts are imposed the AS peak azimuth width may be reduced, and a 1D Gaussian then becomes the preferred AS 1D peak model.

Figure~\ref{fourierfig} (right panel) shows the SS 2D peak from Fig.~\ref{isolate} (left panel) projected onto $\phi_\Delta$ (bold solid  histogram). The fitted SS 1D Gaussian (dash-dotted curve) and AS dipole (dashed curve) are also shown. The Fourier components (azimuth multipoles) of the SS peak obtained from Eq.~(\ref{fourier}) are represented by bold dotted curves with amplitudes 
$A_X\{SS\}(b) = 2\rho_0(b) v_m^2\{SS\}(b) = F_m(\sigma_{\phi_\Delta}) A_{1,proj}(b)$, 
where $A_{1,proj}$ is the amplitude of the SS 2D peak when projected onto 1D azimuth and letters $X = D,\, Q,\, S,\, O$ denote cylindrical azimuth multipoles: dipole, quadrupole, sextupole, octupole, with pole number $2m$.  The standard fit model of Eq.~(\ref{estructfit}) provides an accurate description of 2D angular correlations, their 1D projections and any multipoles derived from 1D Fourier analysis of the SS 2D peak.

\section{Imposing a 1D Fourier series on $\bf \phi_\Delta$} \label{imposing}

This section establishes context for the main subject of this paper  presented in Sec.~\ref{2dsex}. Fourier series analysis applied to projections of all 2D angular correlations onto 1D azimuth has been coupled with interpretation of each sinusoid term as a flow manifestation. But Fourier series derived from 1D projections cannot describe 2D angular correlations accurately, and  1D projections can be represented more efficiently with alternative peaked model functions. Several examples are considered below.

\subsection{``Elliptic flow'' measurements}

The first report of $v_2$ measurements (azimuth quadrupole) derived from RHIC collisions is Ref.~\cite{v2one} where the ``event-plane'' (EP) method was invoked. We have demonstrated that such nongraphical numerical methods (EP, two-particle cumulants) are equivalent to fitting the sum of all angular correlations projected onto 1D $\phi_\Delta$ with a single sinusoid $\cos(2\phi_\Delta)$~\cite{flowmeth}. Thus, such $v_2$ results {\em must} include contributions from the jet-related SS 2D peak~\cite{gluequad,multipoles}. Several schemes have been proposed to distinguish ``flow'' from ``nonflow'' (i.e., some fraction of the jet contribution)~\cite{2004}. But alternative $v_2$ methods typically rely on strong assumptions about the properties of statistical measures (e.g., $v_2\{4\}$) that are actually contradicted by the data~\cite{davidhq,davidhq2}. And assumptions about how ``nonflow'' (jet structure) is distributed on $\eta$ can also be questioned. The consequence may be substantial jet-related bias in published $v_2$ results~\cite{davidhq2,multipoles}.

\subsection{Dihadron correlations and ZYAM subtraction}

Dihadron correlations on 1D azimuth with various trigger-associated $p_t$ cuts are expected to reveal jet structure in more-central \aa collisions where event-wise jet reconstruction is difficult. Basic (raw) dihadron correlations include a large combinatoric background. Attempts to isolate jet-related correlation structure typically rely on subtracting an estimate of that background~\cite{dihadron}. 

The background estimation is based on some criterion for determining a constant offset value and a quadrupole modulation amplitude derived from published $v_2$ data. The conventional method is denoted ZYAM (zero yield at minimum) subtraction. If $v_2$ data are biased (e.g., by some fraction of the SS 2D peak quadrupole component), and/or the ZYAM offset criterion is invalid (e.g., SS and AS peaks overlap significantly), the background estimate may be very inaccurate, leading to underestimation and distortion of inferred jet structure~\cite{tzyam}. 

\subsection{``Triangular flow'' on 1D $\phi_\Delta$}

One of the consequences of ZYAM subtraction is the emergence of a double peak in the resulting AS correlation structure in place of the expected single broad AS peak (AS dipole) corresponding to back-to-back jet correlations~\cite{dihadron}. The unexpected AS structure led to theoretical conjectures about formation of Mach cones by interactions of energetic partons with a dense medium~\cite{mach}. 

More recently it was proposed that such AS structure may be a final-state manifestation of  ``triangular flow'' arising from corresponding structure (``triangularity'') in the initial-state \aa geometry, via radial expansion of a bulk medium~\cite{gunther}. The underlying assumption of such triangular flow analysis is that all correlation structure projected onto 1D azimuth represents flows described by the terms of a 1D Fourier series representation. But {\em any} distribution on periodic azimuth can be described accurately by a 1D Fourier series, and there is no obligation to interpret the Fourier terms as representing flows.

\subsection{The $\eta$ dependence of ``triangular flow''}

Based on the assumption that  inference of triangular flow from 1D Fourier analysis of projected 2D angular correlations is a valid concept it is proposed to measure the $\eta$ dependence of triangular flow by binning the difference variable $\eta_\Delta$ and representing the projected azimuth correlations in each \deta bin by a Fourier series, with the $m=3$ sextupole term emphasized~\cite{sorentriang}. We apply that method to a particular case (centrality) with direct comparison to standard 2D correlation analysis.  In principle the same argument should apply to any \aa centrality.

 \begin{figure}[h]
 \includegraphics[width=1.65in,height=1.6in]{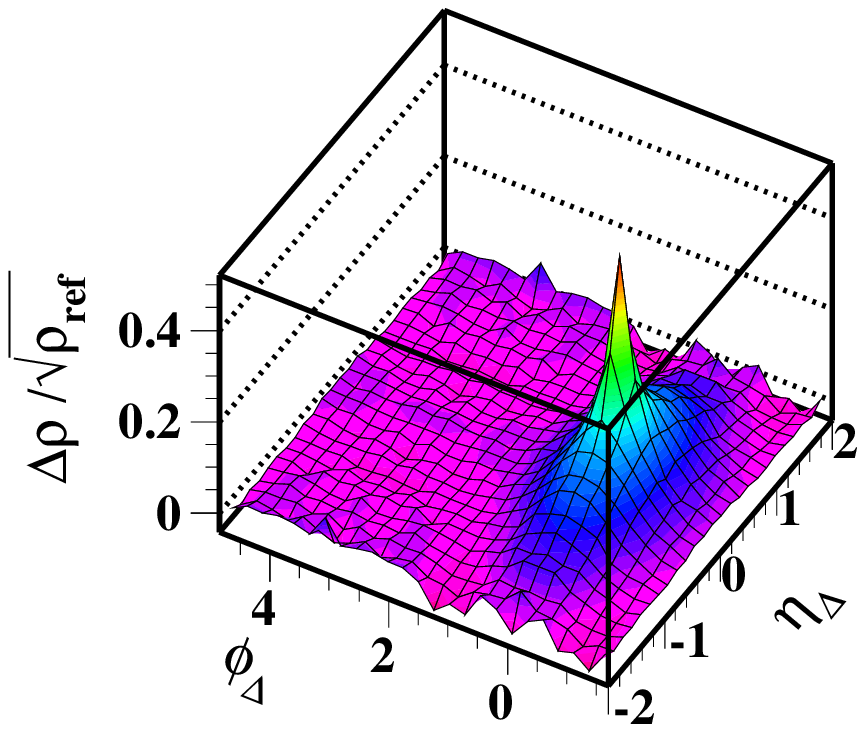}
 \includegraphics[width=1.65in,height=1.6in]{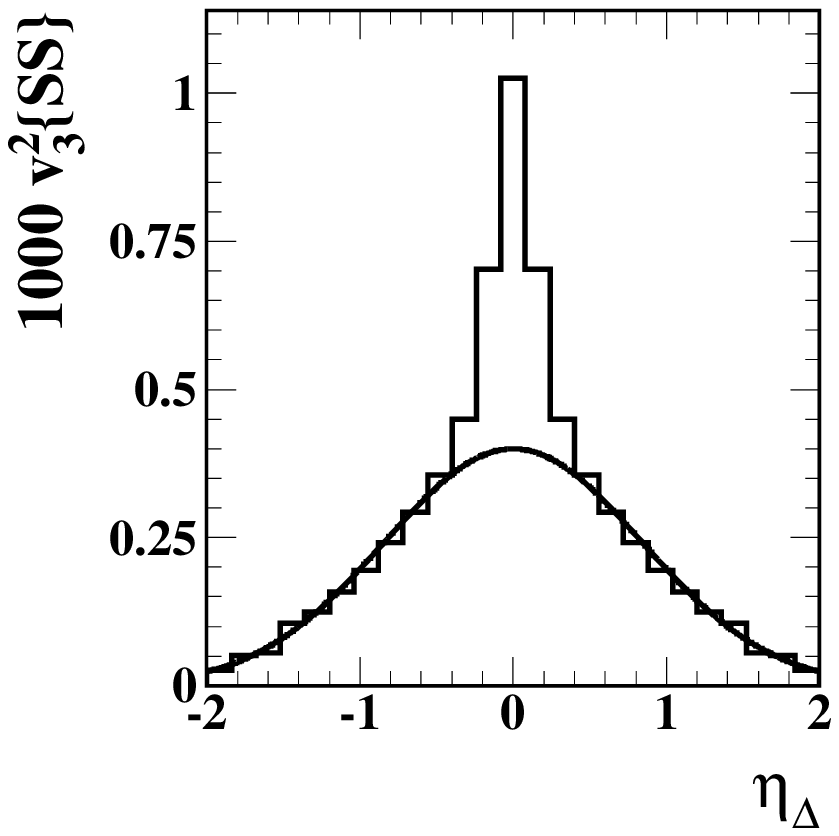}
\caption{\label{triangdata}
Left: Data from a $p_t$-integral analysis of 46-55\%-central 200 GeV \auau collisions from which fitted AS dipole and nonjet quadrupole (plus constant offset) components have been subtracted, leaving the SS 2D peak. The narrow peak at the origin is Bose-Einstein correlations (BEC) plus conversion-electron pairs.
Right: The SS azimuth sextupole component of the data in the left panel for each $\eta_\Delta$ bin. The solid curve is derived from the fitted SS 2D (jet) peak model.
} 
 \end{figure}

Figure~\ref{triangdata} (left panel) shows the SS 2D peak obtained from 46-55\%-central 200 GeV \auau collisions~\cite{anomalous}. The standard fit model consisting of AS dipole plus NJ quadrupole plus SS 2D peak provides an excellent data description. In the left panel the fitted offset, AS dipole and nonjet quadrupole have been subtracted from the 2D histogram leaving the only possible source of ``higher harmonics'' with $m > 2$. That centrality bin lies just below the ``sharp transition'' in angular correlations, in the interval where jet-related structure strictly follows the GLS scaling of  \nn jet production expected for ``transparent'' \aa collisions (see Fig.~\ref{centdep}). The SS peak volume is predicted by pQCD binary-collision scaling of dijet production~\cite{jetsyield}. Thus, a jet interpretation for the SS 2D peak is well supported.

It is straightforward to convert measured SS 2D peak properties to $v_3^2\{SS\}(\eta_\Delta)$. First we separate the fitted narrow BEC-electron peak from the underlying broader SS 2D jet peak with their different azimuth widths. We convert the latter to the broader $v_3^2\{SS\}(\eta_\Delta)$ distribution on $\eta_\Delta$ using the measured 2D peak properties and Eq.~(\ref{fourier}), especially the fixed azimuth width. We do the same with the narrower fitted BEC-electron peak (with its much smaller azimuth width) and recombine the two. 

Figure~\ref{triangdata} (right panel) shows the result. The broader underlying peak is a Fourier component of the jet-related SS 2D data peak. The solid curve (1D Gaussian) is derived from the 2D model fit. The BEC-electron peak is proportionately much larger compared to the jet peak in this case. The BEC peak is narrower on azimuth, so the Fourier components for greater $m$ are relatively larger, consistent with Eq.~(\ref{fourier}).
This exercise demonstrates that previously-measured 2D angular correlations, including jet-related structure, can predict any ``triangular flow'' result obtained as a function of $\eta_\Delta$ with a non-graphical numerical method. Any such Fourier terms with $m > 2$ must be manifestations of the monolithic SS 2D jet peak.

\subsection{Lower and higher ``harmonic flows''}

\begin{figure}[h]
 \includegraphics[width=1.65in,height=1.6in]{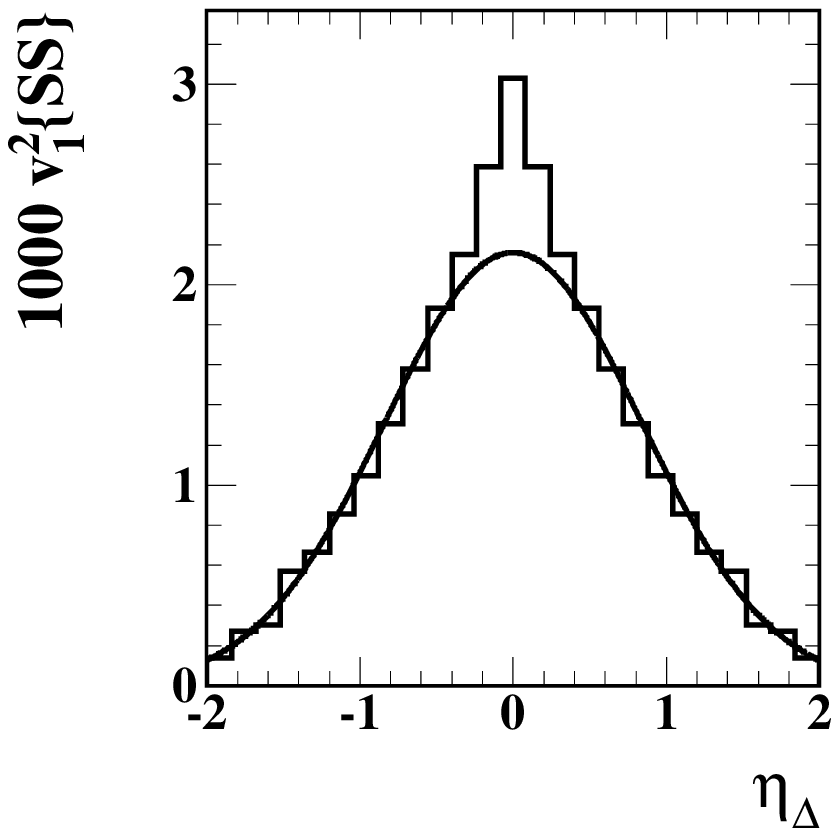}
 \includegraphics[width=1.65in,height=1.6in]{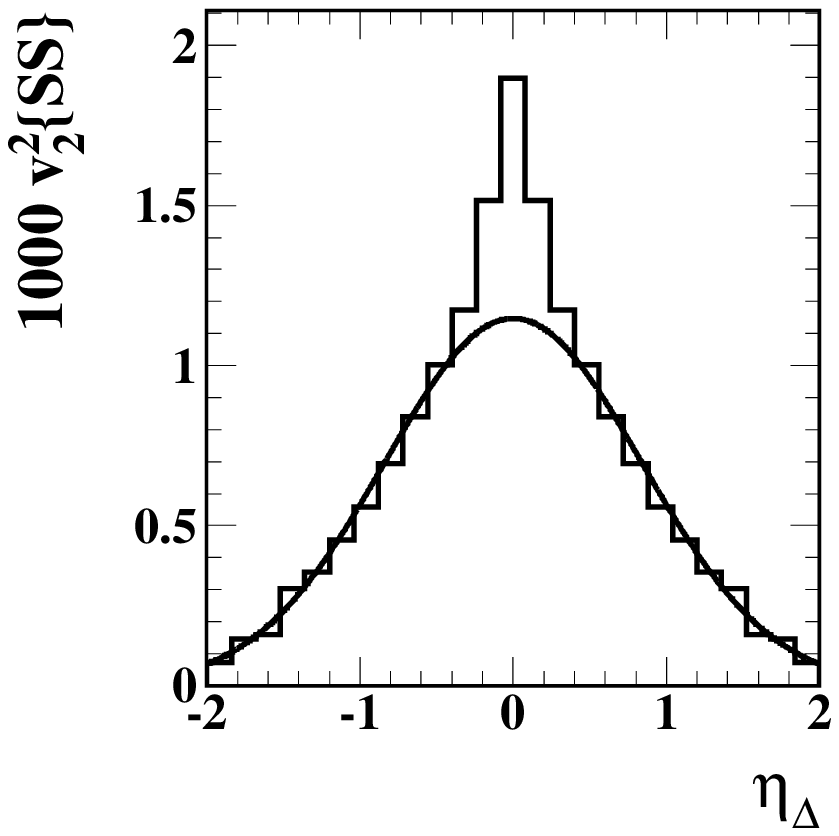}
\caption{\label{highlow}
Left: The SS azimuth dipole component of the data  from Fig.~\ref{triangdata} (left panel) for each $\eta_\Delta$ bin.
Right: The SS azimuth quadrupole component of the same data. The solid curves are derived from the fitted SS 2D peak model.
} 
 \end{figure}

The same procedure can be carried out with lower-order ($m < 3$) multipoles as well. Figure~\ref{highlow} illustrates the procedure applied to the SS 2D peak in Fig.~\ref{triangdata} (left panel) for $m  =1$ (SS dipole) and 2 (SS quadrupole). The relative heights of jet-related and BEC-electron peaks depend on the different azimuth widths according to Eq.~(\ref{fourier}). The right panel shows the dominant (jet-related) source of ``nonflow'' bias in conventional $v_2$ data. The hallmark of such SS 2D peak contributions is a large curvature on $\eta_\Delta$, in contrast to the approximate uniformity on $\eta_\Delta$ of the nonjet (NJ) quadrupole within the STAR TPC acceptance.

Higher harmonic flows ($m > 2$) inferred from 1D projections onto azimuth have also been obtained from LHC data~\cite{alice,atlas}. In those studies imposed ``$\eta$ exclusion cuts'' are intended to exclude jet contributions {\em assumed} to be restricted to an $\eta$ interval near the origin. But the definition of jet structure in more-central \aa collisions is a major question as yet unresolved. In such analysis the presence of the sharp transition in SS 2D peak properties is still evident at LHC energies. In Ref.~\cite{multipoles} it is demonstrated that all trends associated with ``higher harmonic flows'' are predicted by jet-related angular correlations. The same multipole trends observed at 62 and 200 GeV are also evident at 2.76 TeV. The only difference is a common extrapolation factor 1.3 applied to all such trends.

\section{A sextupole in the 2D  fit model} \label{2dsex}

We now confront the central question for this study. We have demonstrated that any ``higher harmonics'' inferred from 1D Fourier analysis applied to projections of 2D angular correlations onto azimuth are at least dominated by a contribution from the SS 2D peak, and SS peak properties are consistent with jet production in at least some cases. We now consider addition of a separate sextupole element to the standard 2D fit model motivated by a desire for increased sensitivity to ``triangular flow.'' If we observe a significant nonzero sextupole amplitude does that imply a nonjet sextupole component in the data that might indeed represent triangular flow?

To answer that question we compare three 2D model configurations applied to the same data histogram: (a) the standard 2D model, (b) the standard fit model plus SS sextupole element, and (c) the standard fit model plus SS 1D Gaussian on $\phi_\Delta$. The relevance of configuration (c) will become apparent below. With $p_t$-integral data we show that if a sextupole element uniform on $\eta_\Delta$ is added to the standard 2D fit model nonzero sextupole amplitudes are indeed inferred from model fits to more-central Au-Au data. But other model parameters are also shifted substantially from their original values. The {\em combination} of changes is equivalent to a single new model element. The added sextupole element does not reveal an independent nonjet sextupole component in the data. 

\subsection{Direct comparison of three 2D fit models} \label{dircomp}

Table~\ref{params} shows parameters from three 2D fits to the same data histogram from 9-18\% central \auau collisions. Other more-central \auau data give similar results. Fit parameters that do not change significantly between models are not relevant to this discussion and are omitted from the Table. The parameter labels are as defined in Eq.~(\ref{estructfit}).

\begin{table}[h]
  \caption{Model parameters for three fit models: (a)  standard model for this analysis, (b) standard model plus additional sextupole term $A_S$ and (c)  standard model plus additional SS 1D Gaussian on $\phi_\Delta$ with amplitude $A_{1D}$ as part of the SS 2D peak model. The model parameters are as defined in Eq.~(\ref{estructfit}). Uncertainties in the second column illustrate typical fit uncertainties (statistical plus systematic) for each parameter. $A_N$ denotes the additional (new) model parameter -- $A_S$ or $A_{1D}$ in third and fourth columns respectively.
}
  \label{params}
\begin{center}
\begin{tabular}{|c|c|c|c|} \hline
 parameter &standard & std + $A_S$ & std + $A_{1D}$   \\ \hline
$A_{1}$  & $0.76 \pm 0.04$ & 0.51 & 0.47  \\ \hline
 $\sigma_{\eta_\Delta}$ & 2.3$\pm 0.3$ & 1.78 & 1.72  \\ \hline
$A_Q$ &0.18$\pm 0.01$ & 0.23 & 0.18   \\ \hline
$A_D$ &0.29$\pm 0.02$ & 0.175 & 0.28   \\ \hline
$A_{N}$ & -- & 0.012 & 0.29   \\ \hline
\end{tabular}
\end{center}
\end{table}

The second column (standard) shows fit results obtained with the standard 2D model. Those values are consistent within fit uncertainties with the results in Table III of Ref.~\cite{anomalous}. The third column (std + $A_S$) shows fit results when a sextupole term $A_S \cos(3 \phi_\Delta)$ is added to the standard model and all else remains the same.  The fourth column shows fit results when a SS 1D Gaussian on azimuth $A_{1D}\exp[-(\phi_\Delta / \sigma_{\phi_\Delta,1D})^2/2]$ is added to the standard fit model and all else remains the same.

\subsection{Model equivalence: sextupole vs 1D Gaussian} \label{equiv}

We now demonstrate that addition of a sextupole term $A_S$ to the standard model is equivalent to modifying the SS 2D peak model by adding a SS 1D Gaussian narrow on $\phi_\Delta$ and uniform on $\eta_\Delta$.  The argument is based on results presented in Refs.~\cite{tzyam,multipoles} where periodic peak arrays are approximated by truncated Fourier series.  

The model elements $A_D$ (dipole) and $A_Q$ (quadrupole) already present in the standard 2D fit model are in effect parts of a (truncated) Fourier series. Adding an $A_S$ (sextupole) term to the data model can extend the Fourier series to $m=3$. The next term $A_O$ (octupole) is typically at the level of statistical uncertainties~\cite{multipoles,gunther}. The separate sinusoids may serve {\em both} as representatives of distinct (nonjet) data multipoles and as parts of a Fourier series representing a localized (possibly jet-related) peak structure on azimuth~\cite{tzyam,multipoles}.

\begin{table}[h]
  \caption{Comparison of {\em changes} in multipole amplitudes resulting from addition of a sextupole element to the standard fit model with Fourier coefficients of a SS 1D Gaussian on azimuth. The comparison demonstrates equivalence between the changes in multipole amplitudes (columns 2 and 3 of Table~\ref{params}) and the Fourier coefficients of a 1D Gaussian on azimuth (column 5 of this Table). Column 5 is obtained by multiplying Fourier coefficients $F_m$ by the value $A_{1D} = 0.25$ that best matches the third column. That value corresponds to the difference $\Delta A_{1} = 0.76-0.51$ from Table~\ref{params}.}
  \label{fourierr}
\begin{center}
\begin{tabular}{|c|c|c|c|c|} \hline
 $m$ &parameter & data & $F_m$ & $F_m\, A_{1D}$   \\ \hline
1 &$-\Delta A_D$ &0.115 & 0.445 & 0.11   \\ \hline
2 & $\Delta A_Q$ &0.05 & 0.20 & 0.05   \\ \hline
3 & $\Delta A_S$ & 0.012 & 0.053 & 0.013   \\ \hline
\end{tabular}
\end{center}
\end{table}

Table~\ref{fourierr} (third column, data) shows the {\em changes} in fitted multipole amplitudes between the standard model and that including the sextupole term. The fourth column, with $F_m$ defined in Eq.~(\ref{fourier}), shows the calculated Fourier coefficients for a unit-amplitude 1D Gaussian on azimuth with width $\sigma_{\phi_\Delta,1D} = 0.73$~\cite{multipoles}. The fifth column ($F_m A_{1D}$) shows the Fourier coefficients for a SS 1D Gaussian with amplitude $A_{1D} = 0.25$. The Gaussian amplitude and width were adjusted to achieve the best match between columns 3 and 5, achieving equivalence within small data uncertainties. 
The amplitude $A_{1D}$ of the added SS 1D Gaussian on azimuth corresponds to the difference $\Delta A_1$ between SS 2D peak amplitudes in Table~\ref{params} for the two fit models. And the inferred 1D peak width approximates the azimuth width of the SS 2D peak in the standard fit model. Thus, the parameter changes resulting from inclusion of a sextupole element in the 2D model are equivalent to adding a constant offset $A_{1D}$ to the $H(\eta_\Delta)$ factor of the SS 2D peak model.

Table~\ref{params} (fourth column) confirms the equivalence with a 2D fit that replaces the added sextupole element by an offset in the $\eta_\Delta$ factor of the SS 2D peak model [$A_1\exp(-\eta_\Delta^2 / 2 \sigma^2_{\eta_\Delta}) \rightarrow A'_1\exp(-\eta_\Delta^2 / 2 \sigma^{\prime2}_{\eta_\Delta}) + A_{1D}$]. The added constant in $H(\eta_\Delta)$ then represents a SS 1D Gaussian on $\phi_\Delta$ in the 2D model [see Eq.~(\ref{estructfit})]. 
The fit results are consistent with the previous exercise: the fitted 1D Gaussian amplitude is $A_{1D} = 0.29$ compared to 0.25, and the modified 2D amplitude $A'_1 = 0.47$ is consistent with $A_{1D}$ (i.e., $A'_1 + A_{1D} = A_1 = 0.76$).  
Note that in column 4 $A_D$ and $A_Q$ {\em have reverted to their values for the standard fit model}, since the added SS 1D Gaussian on azimuth is modeled explicitly by a Gaussian function rather than by a truncated Fourier series.

 \begin{figure}[h]
 \includegraphics[width=3.3in,height=1.6in]{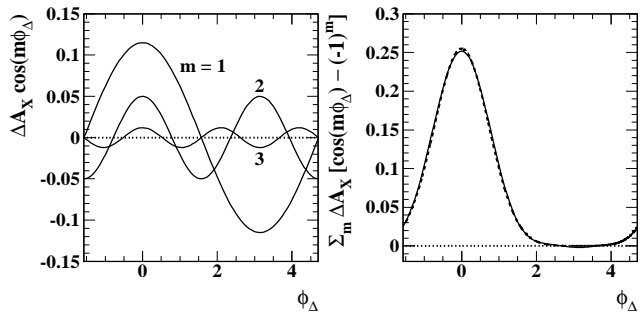}
\caption{\label{sex1d}
Left: The curves represent the {\em changes} in three Fourier amplitudes (multipoles) between the standard fit model and a model including an added sextupole element applied to the same 2D data histogram, as reported in Table~\ref{fourierr}. Subscript $X$ represents dipole D, quadrupole Q and sextupole S for $m = 1$, 2 and 3 respectively.
Right: The sum of the three sinusoids in the left panel per the axis label (solid curve) compared to a {\em periodic} 1D Gaussian (dashed curve).
} 
 \end{figure}

Figure~\ref{sex1d} illustrates the equivalence graphically. The left panel shows the {\em changes} in three multipoles, between the standard 2D model and that with the added sextupole element. The sinusoid amplitudes are as in Table~\ref{fourierr} (third column). The right panel shows (solid curve) the sum of the three sinusoids in the left panel. The dashed curve is a SS 1D Gaussian with amplitude 0.255 and width 0.73 demonstrating the accurate equivalence. Remaining small differences are consistent with the missing octupole term ($m=4$) in the truncated Fourier series.

\subsection{Model ambiguities and parameter significance}

The previous exercise demonstrated that the nonzero sextupole amplitude resulting from fits with 2D model (b) in Table~\ref{params} is actually associated with the jet-related SS 2D peak. We now return to the {\em significance} of 2D model-fit results. As noted in Sec.~\ref{factorize} the standard fit model of the SS 2D peak in Eq.~(\ref{estructfit}) is effectively factorized to 1D functions on $\eta_\Delta$ and $\phi_\Delta$. Whatever the actual peak structure $H(\eta_\Delta)$ we do observe that the narrow 1D Gaussian $G(\phi_\Delta)$ is independent of $\eta_\Delta$, and factorization accurately describes the SS 2D peak. We can therefore simplify the model comparison in Sec.~\ref{dircomp} to a single 1D Gaussian on $\eta_\Delta$ with parameters $(A,\sigma_\eta)$ vs a 1D Gaussian plus constant offset with parameters $(A',\sigma'_\eta,A_{1D})$.

\begin{figure}[h]  
\includegraphics[width=.23\textwidth,height=.23\textwidth]{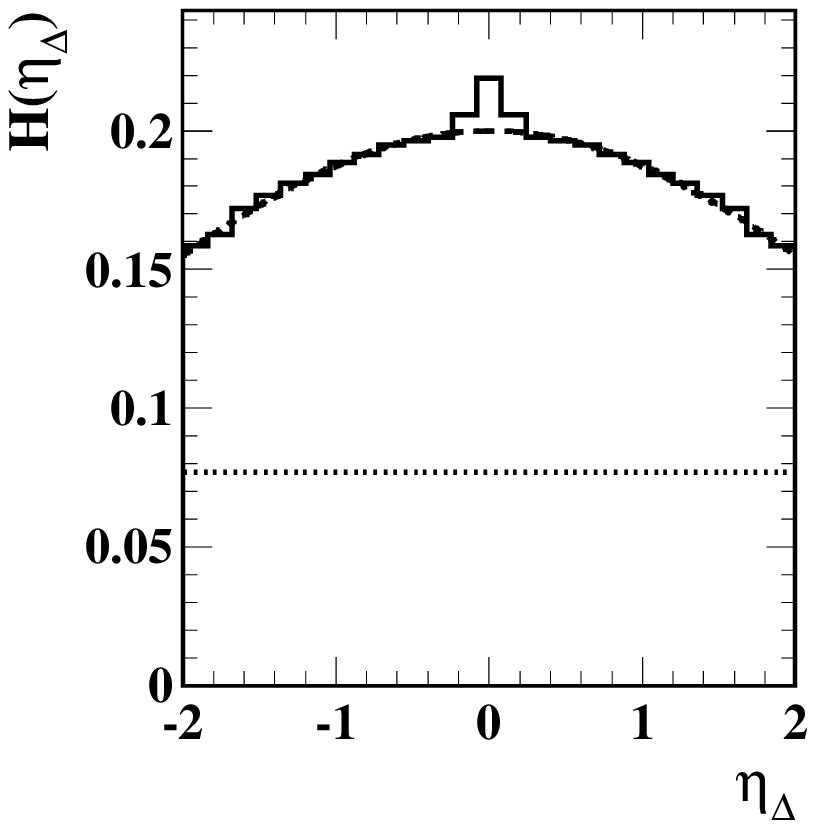}  
\includegraphics[width=.23\textwidth,height=.235\textwidth]{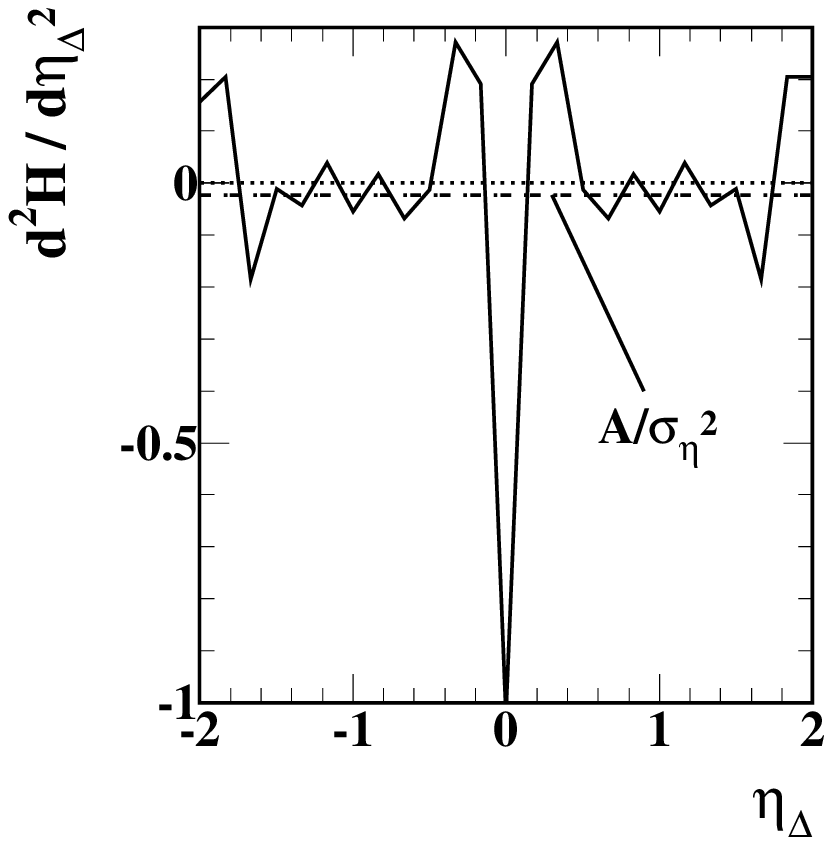}  
\caption{\label{modell}
Left: 
The data histogram is the 2D histogram in Fig.~\ref{isolate} (left panel) projected onto 1D $\eta_\Delta$. The dashed curve through the data histogram is a single 1D Gaussian with parameters $(A,\sigma_\eta)$. The dash-dotted curve is the sum of a 1D Gaussian $(A',\sigma'_\eta)$ plus a constant offset $A_{1D}$ represented by the dotted line. The two curves cannot be resolved within the $\eta$ acceptance limits.
Right:  The second difference of the data histogram in the left panel corresponds to its local curvature. The calculated curvature $A / \sigma^2_\eta$ of the 1D Gaussian in the left panel (dashed curve) is the dash-dotted line in this panel. 
}  
\end{figure}

Figure~\ref{modell} (left panel) repeats the projected 1D histogram in Fig.~\ref{isolate} (right panel) . 
The projected histogram is accurately modeled  by a 1D Gaussian (dashed curve) with amplitude $A = 0.20$ and width $\sigma_\eta = 2.7$. The 1D peak symmetric about $\eta_\Delta = 0$ can also be described in a {\em model-independent} way by a Taylor series in even powers of $\eta_\Delta$. Two terms are systematically significant within the $\eta$ acceptance: a constant term (peak amplitude) and a quadratic term (peak curvature at the mode). The quartic term is not statistically significant. 

In the right panel the Gaussian curvature value $A / \sigma_{\eta}^2 = -0.027 $ (dash-dotted line) is plotted together with the bin-wise second difference of the data histogram, comparing local  statistical noise to the curvature signal. The fourth difference (quartic) would be overwhelmed by statistical fluctuations. The Gaussian curvature is consistent with the Taylor series.  The number of model parameters thus matches the data structure (significant series coefficients) and the data are accurately described.

If a constant offset is added to the 1D fit model the data can constrain only certain {\em combinations} of the parameters. The constraints are $A = A' + A_{1D}$ for the common peak amplitude and $A/\sigma_\eta^2 = A'/\sigma^{\prime2}_\eta$ for the common peak curvature at the mode. The parameters  $(A',\sigma'_\eta,A_{1D})$ are otherwise {\em free to vary} within those constraints. If we choose the value $A_{1D} = 0.077$ (dotted line in the left panel, equivalent to $A_{1D} = 0.29$ for the unprojected 2D peak model) the data constraints determine that $A' = 0.123$ and $\sigma'_\eta = 2.12$. A dash-dotted curve representing the sum of the modified 1D Gaussian and the constant offset is superposed in the left panel upon the dashed Gaussian curve. The curves are not distinguishable {\em within the $\eta$ acceptance limits}. 

The same exercise could be carried out with a range of $A_{1D}$ values {\em including zero} (the standard data model). Neither the offset parameter $A_{1D}$ nor the mathematically equivalent sextupole amplitude $A_S$ is {\em systematically significant} for data within the STAR TPC acceptance. Changes in fitted $A_S$, $A_D$ and $A_Q$ are strongly correlated and must be combined to determine the actual changes in the 2D model. In the present case the {\em effective} changes in the model are negligible, even though some individual parameters may vary substantially. Considering individual model parameters in isolation (e.g.,  ignoring strong covariances) can be very misleading.

 \section{Nonjet ``Higher harmonics''}  \label{higherharm}


The (jet-related) SS 2D peak {\em must} contribute to  higher multipoles inferred from 1D Fourier analysis. And we demonstrated that the SS peak can also contribute an apparent jet-related sextupole for 2D model fits that are insufficiently constrained by data. But can it be demonstrated that there is a significant contribution to higher multipoles {\em not} associated with the SS 2D peak? In other words, are {\em nonjet} higher multipoles {\em necessary} to describe measured 2D angular correlations? We must establish what nonjet multipole elements are necessary to describe 2D angular correlation data, and what elements are not required by the data or are excluded.

\subsection{Necessity of the nonjet quadrupole element}

 \begin{figure}[h]
 \includegraphics[width=1.65in]{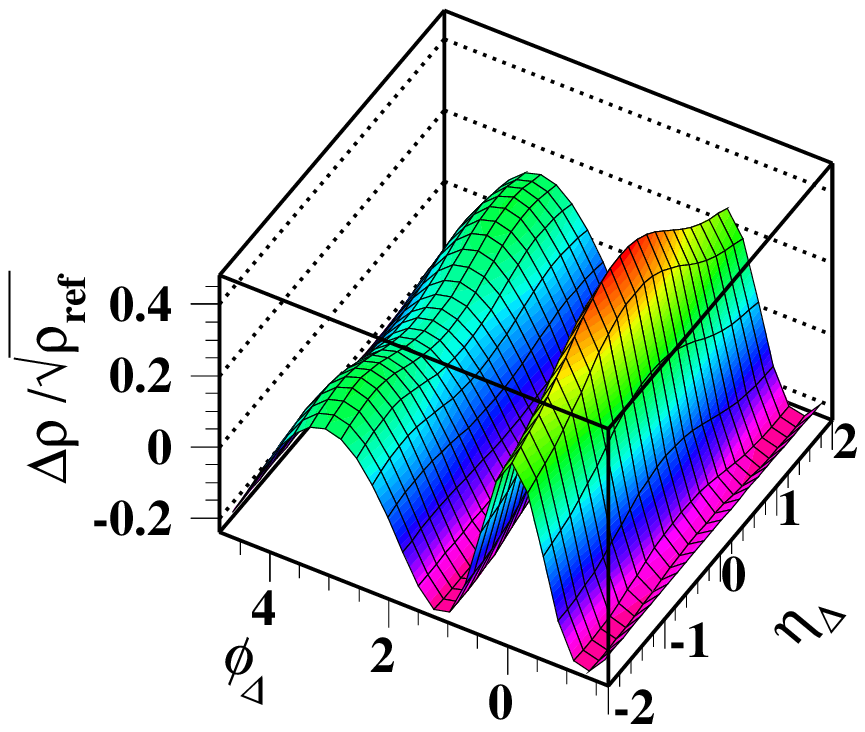}
\put(-100,80){(a)}
 \includegraphics[width=1.65in]{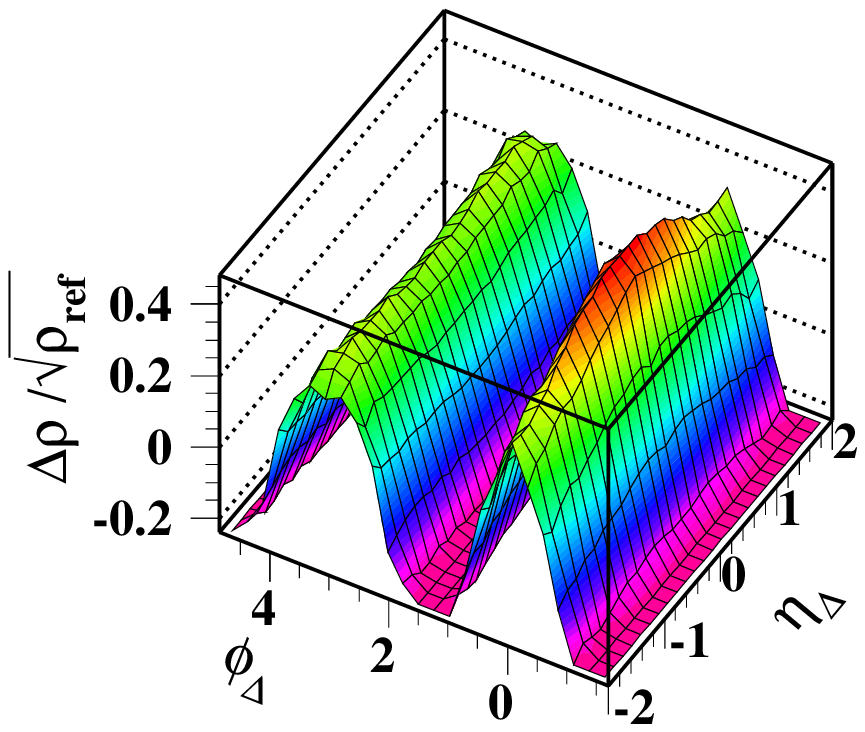}
 \put(-100,80){(b)}\\
\includegraphics[width=1.65in]{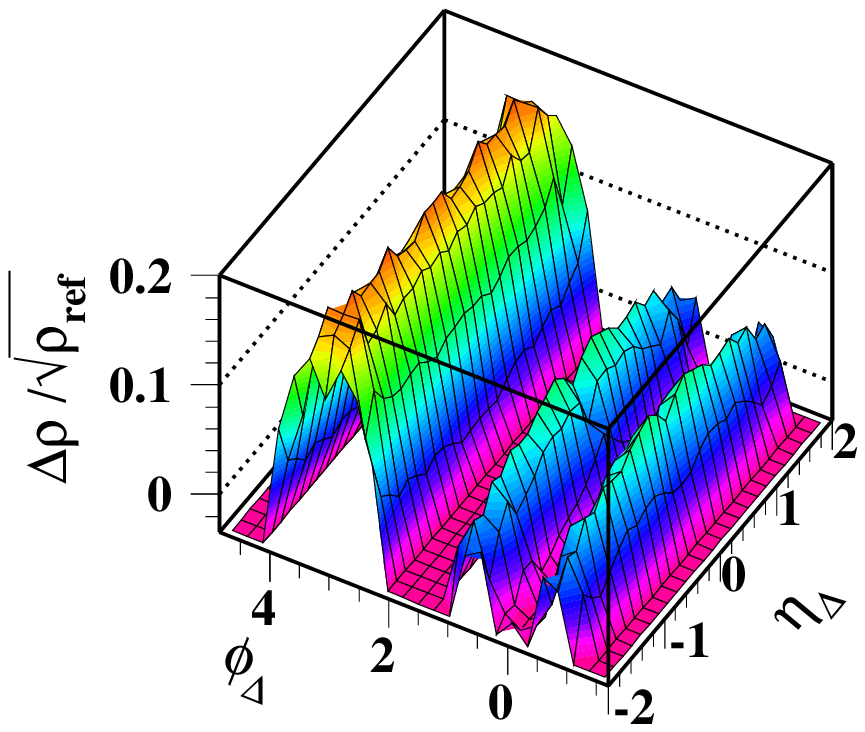}
\put(-100,80){(c)}
 \includegraphics[width=1.65in]{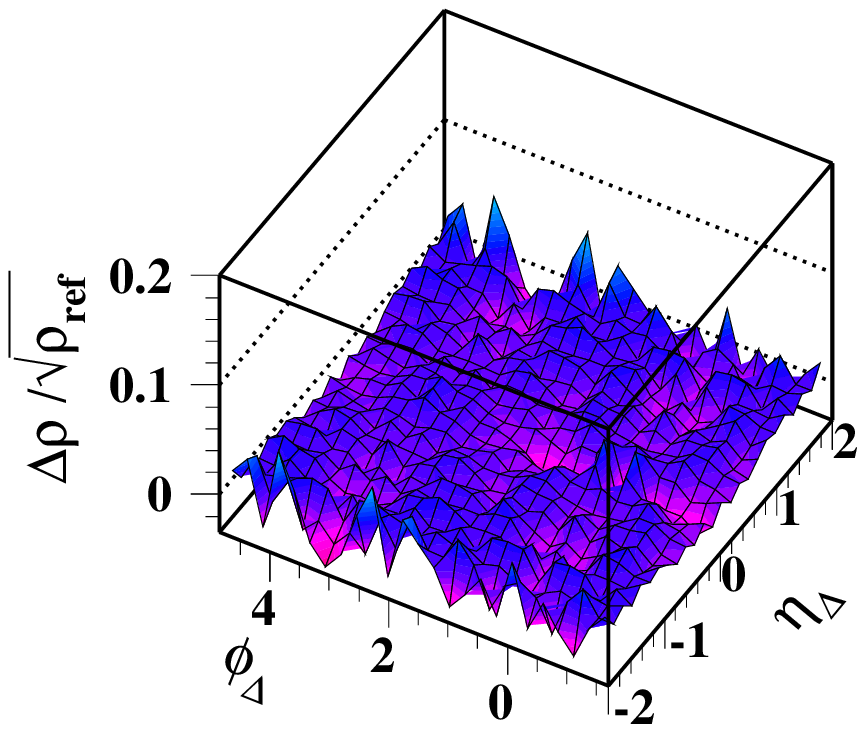}
\put(-100,80){(d)}
\caption{\label{quadnecc} Model fit to  $p_t$-integral data from 9-18\%-central 200 GeV \auau collisions demonstrating the necessity of a nonjet quadrupole model element. (a) fitted model without quadrupole element, (b) data (c) fit residuals without nonjet quadrupole element, (d) residuals with standard model fit.
} 
 \end{figure}

Figure~\ref{quadnecc} provides a demonstration that the nonjet {\em quadrupole} is a necessary element of the standard 2D fit model. The $p_t$-integral data are from 9-18\%-central \auau collisions. The standard fit model provides an excellent description of those data, and a nonjet quadrupole amplitude $A_Q = 0.18 \pm 0.01$ is inferred for that centrality~\cite{anomalous}. The fit residuals for the standard fit model shown in panel (d) are consistent with statistical uncertainties. The other panels reveal the consequences of omitting the nonjet quadrupole element from the fit model. 

The best-fit model in panel (a) does not include a nonjet quadrupole element. The corresponding fit residuals in panel (c) reveal at least 80\% of the correct $A_Q$ amplitude near $\phi_\Delta = \pi$. To minimize $\chi^2$ with the improper model the fitting routine has compensated by overestimating the SS 2D peak amplitude by about 30\% (0.76 $\rightarrow$ 1.0), leading to the SS minimum in the residuals at $\phi_\Delta = 0$. The fit residuals in panel (c) demonstrate that the nonjet quadrupole, a quadrupole component independent of the SS 2D peak, is a {necessary} model element. 

\subsection{Adding a nonjet sextupole component to data} \label{nononjet}

\begin{figure}[h]
 \includegraphics[width=1.65in]{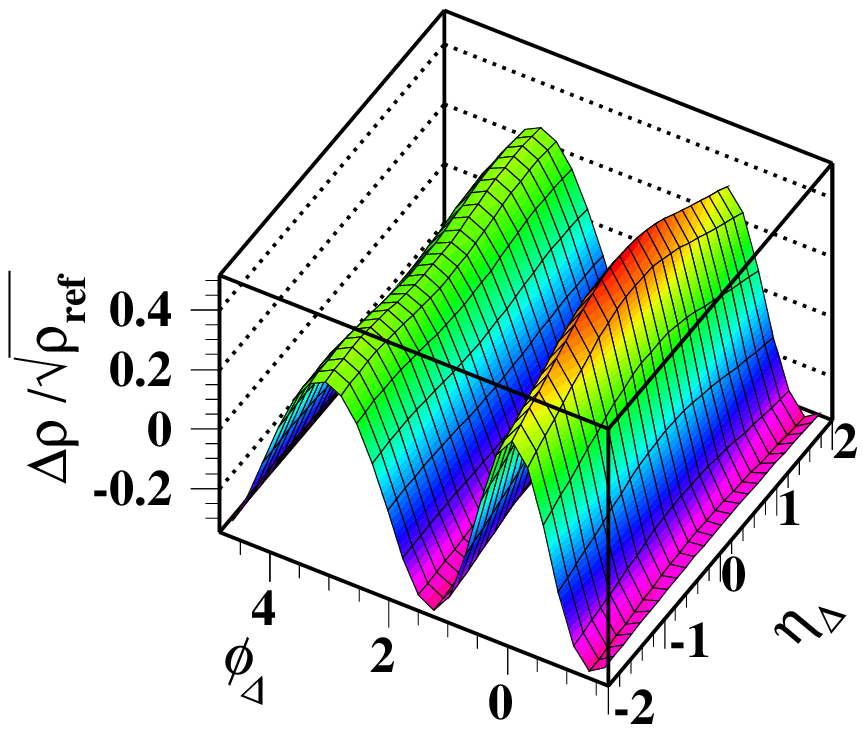}
\put(-100,80){(a)}
 \includegraphics[width=1.65in]{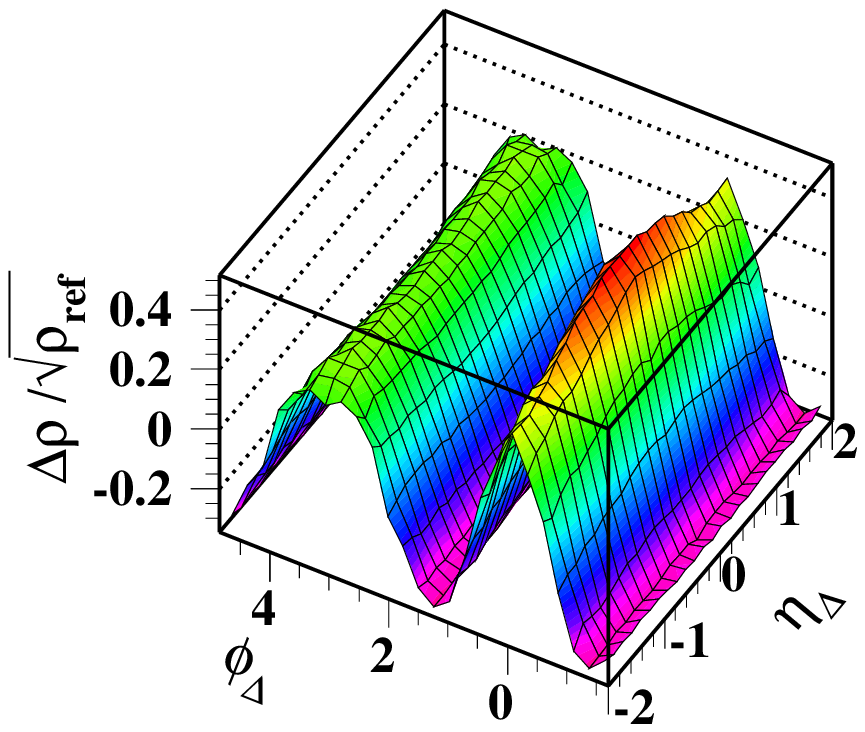}
\put(-100,80){(b)}\\
 \includegraphics[width=1.65in]{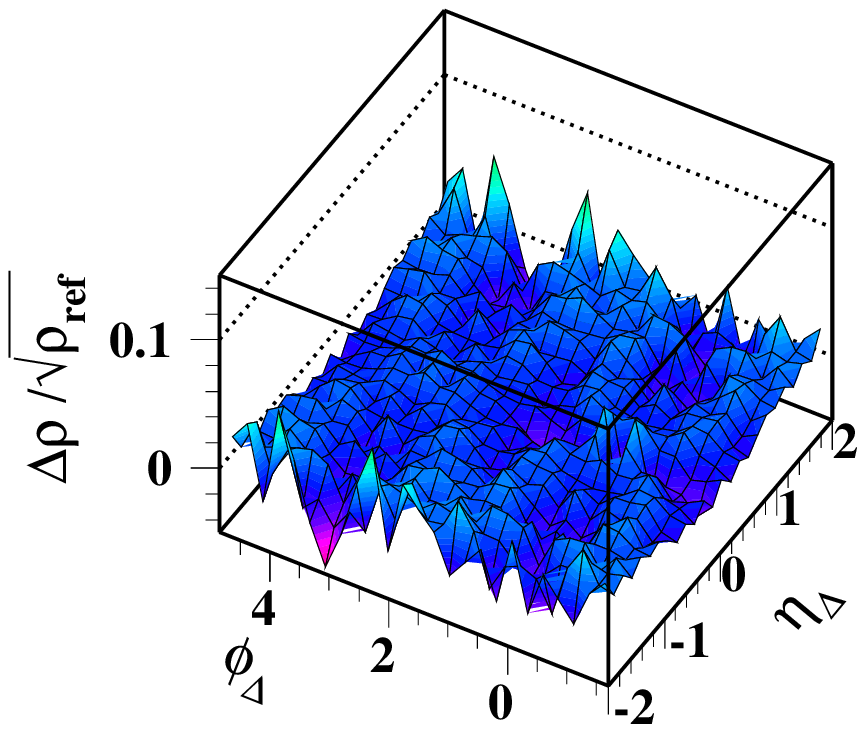}
\put(-100,80){(c)}
 \includegraphics[width=1.65in]{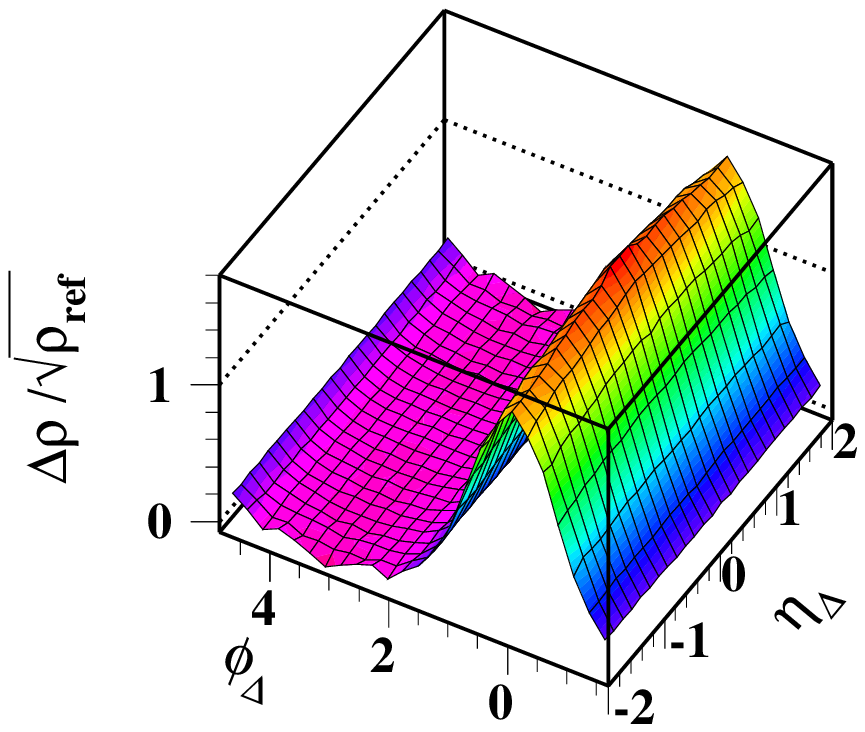}
\put(-100,80){(d)}
\caption{\label{triangnecc}
Model fit to  $p_t$-integral data from 9-18\%-central 200 GeV \auau collisions to which a {\em nonjet} sextupole component has been added. (a) Standard fit model, (b) data with sextupole component added (c) fit residuals, (d) corresponding inferred SS 2D peak. 
} 
 \end{figure}

Figure~\ref{triangnecc} explores the same question for higher multipoles: Can we establish the necessity of a nonjet sextupole from data analysis.
We determine the sensitivity of the fitting process to a nonjet sextupole by deliberately adding such a term to the 2D data and refitting with the standard 2D model having no nonjet sextupole element. 

In Fig.~\ref{triangnecc}  a sextupole component has been added to the input data with $A_S\{2D\} = 0.034$. The reason for that value is explained below. Panel (a) shows the standard model fit, panel (b) shows the modified input data, and panel (c) shows the resulting fit residuals, with four-times increased sensitivity.  The standard-model elements do minimize $\chi^2$ by reducing the {apparent} sextupole component well below what was inserted. The compensation process is essentially a reversal of the process described in Sec.~\ref{equiv}. Changes in the AS dipole (increase) and quadrupole (decrease) interact with a change in the SS 2D Gaussian (increase). Since $A_D$ represents an AS dipole the changes in both dipole and quadrupole are subtracted from the SS peak amplitude increase, leaving a negative SS sextupole component to compensate some fraction of that added to the data. The input value $A_S\{2D\} = 0.034$ chosen for the figure corresponds to reducing the usual $A_Q\{2D\} = 0.18$ value to zero. If a larger sextupole amplitude were introduced the fitted quadrupole value would be driven negative to compensate. If the quadrupole amplitude is constrained to be positive definite a larger input $A_S\{2D\}$ value would result in significant sextupole structure in the residuals.

Panel (d) shows the SS 2D peak inferred by subtracting fitted AS dipole and NJ quadrupole (zero amplitude) model elements from the modified data histogram. Instead of $A_{2D} = 0.76$ inferred from the unmodified data the SS amplitude is now 1.73, and the r.m.s.\ $\eta$ width is 4.3 rather than 2.7. 
Addition of a small sextupole component to the data causes large changes in inferred SS 2D peak structure equivalent to adding a narrow 1D Gaussian with amplitude approximately 1. We conclude  that attempts to isolate {\em nonjet} higher harmonics, multipoles not associated with the SS 2D peak, cannot succeed. 

\section{Discussion} \label{disc}

Conjectured triangular flow and other flow-related multipoles offer the possibility to reinterpret the SS 2D peak in angular correlations as a flow phenomenon rather than a jet manifestation, seeming to buttress claims for formation of a flowing, dense partonic medium with small viscosity in high-energy \aa collisions. In this study we have identified several problems with recent higher-harmonic-flow claims.  We confronted supporting arguments for  higher harmonic flows with observed properties of 2D angular correlation data as reported in Ref.~\cite{anomalous}. 

\subsection{Terminology: jet-related vs nonjet structure}

Proper description and interpretation of 2D angular correlations relies on maintaining distinctions among separate correlation components and model elements. In Sec.~\ref{terminology} we defined ``nonjet'' and ``jet-related'' structures in 2D angular correlations. The terminology is based on data systematics for \pp and more-peripheral \auau collisions where a jet interpretation for those structures is most likely~\cite{anomalous}. We then maintain that language consistently over the full centrality range of \auau collisions. Alternative interpretations of some parts of ``jet-related'' structure are possible in more-central \auau collisions.

In describing multipole elements $v_m$ we distinguish between (a)  $v_m$ obtained from 1D Fourier fits applied to all angular correlations projected onto 1D $\phi_\Delta$ and (b) $v_m$ obtained from fits to angular correlation data with a 2D model that may include one or more multipole elements $A_X$. Multipoles from case (a) must mix jet-related and nonjet structure as discussed in Sec.~\ref{imposing}.  Multipoles from case (b) may represent jet-related or nonjet contributions and are the principal subjects of this study.

\subsection{1D Fourier analysis vs 2D angular correlations}

The possibility to interpret most \auau angular correlation structure in terms of flows has motivated introduction of multiple sinusoids to correlation data modeling. The prototypic example was ``elliptic flow'' analysis based on the ``event-plane''  $v_2\{EP\}$ and two-particle cumulant $v_2\{2\}$ methods~\cite{2004}. Both are equivalent to fitting all angular correlations projected onto 1D $\phi_\Delta$ with a single $\cos(2\phi_\Delta)$ function~\cite{flowmeth,multipoles}. Sinusoid analysis was then extended to include ``triangular flow'' $v_3\{2\}$ by adding a  $\cos(3\phi_\Delta)$ term to 1D model fits~\cite{gunther}. Generalization to a multi-term truncated Fourier series has recently emerged~\cite{luzum,atlas}.
In Sec.~\ref{imposing} we emphasized multipoles inferred by projecting all angular correlations, including the SS 2D peak, onto 1D azimuth and some consequences of isolating individual SS peak Fourier components.

Of course any 1D distribution on periodic azimuth can be represented accurately by a discrete Fourier series---the representation is in some sense trivial. But a 1D Fourier series cannot describe the strong \deta dependence of 2D angular correlations, cannot describe correlations in \pp and more-peripheral \auau collisions that most likely correspond to jet production. And a single 1D series cannot discriminate among several data components that may arise from distinct correlation mechanisms. 

That some data may be described in terms of sinusoids (whatever the fit quality) does not imply that a flow interpretation is justified. The nonjet quadrupole is strongly correlated with initial-state \aa geometry, but its systematics appear to contradict a hydrodynamic interpretation~\cite{davidhq,davidhq2,anomalous}.

The terms of a 1D Fourier series are orthogonal. But orthogonality of Fourier series terms does not imply that a 1D Fourier series is a preferable data model for 2D angular correlations. There are potentially three sources for any sinusoid in a 1D Fourier series: (a) the SS 2D peak, (b) the AS 1D peak, (c) nonjet sinusoids not associated with (a) or (b). The SS 2D peak can be reduced to several azimuth multipoles, but with possibly {\em large curvatures} on $\eta_\Delta$~\cite{multipoles}.  Multiple orthogonal Fourier terms can be strongly correlated by coupling to a single resolved correlation component. And a single Fourier term may be coupled to multiple correlation components and their causative mechanisms (e.g., ``flow'' and ``nonflow'').  

The relation of Fourier series to SS and AS periodic peak arrays is described in Sec.~IV A of Ref.~\cite{tzyam}. Each peak has its own Fourier series. If the two series are combined each term must represent at least two contributions. For instance, the dipole term of a single 1D Fourier series representing all projected 2D angular correlations $A_D\{2\} = A_D\{SS\} - A_D\{AS\}$ is the {\em difference between two large numbers}, the dipole component of the SS 2D peak (positive) and the AS dipole (negative) representing jet structure in at  least some \aa collisions~\cite{anomalous}. 
The small difference has been attributed to ``rapidity-even directed flow'' and ``global momentum conservation''~\cite{directed}. Jet structure is thereby obscured. The net quadrupole term represented by $A_Q\{2\} = A_Q\{SS\} + A_Q\{2D\}$ includes the substantial quadrupole component of the SS 2D peak as well as the nonjet component. The sextupole term is derived from the SS 2D peak as $A_S\{SS\}$.
In contrast, the standard fit model described in Sec.~\ref{stdfit} accommodates all correlation structure on \deta and $\phi_\Delta$. Each model element represents a single observed correlation component, facilitating unique interpretations.

\subsection{Significance of multipoles in 2D data models}

The main problem addressed by this study is the systematic significance of sextupole and higher multipole amplitudes that may be inferred by adding $\cos(m\phi_\Delta)$ terms with $m > 2$ to the standard 2D angular correlation model. If one or more higher multipole elements are added to the standard 2D model are resulting nonzero amplitudes physically meaningful? In this study we have isolated two major issues: (a) additional model elements may exceed what is {\em required} by correlation structure measured within some detector $\eta$ acceptance leading to large systematic uncertainties in certain model parameters, and (b) some {\em fractions} of several model multipoles may combine to form a peaked structure on azimuth that plays an unanticipated role in the 2D data model.

In Sec.~\ref{2dsex} we demonstrated that the SS 2D peak  for data obtained from more-central \auau collisions within a limited $\eta$ acceptance may only constrain two model parameters on $\eta_\Delta$. Addition of a sextupole amplitude to the standard fit model is equivalent to addition of a third parameter (an offset) to the SS peak model. The extraneous  model parameter effectively {\em amplifies} relatively small and systematically insignificant variations in the data to appear as relatively large {\em but still systematically insignificant} variations in the extraneous parameter, but also in other parameters now coupled to it. Although individual model parameters may then seem unstable the {\em overall model} is not changed significantly. The problem is resolved by removing the extraneous parameter(s) per the principle of parsimony or Ockham's razor.

In Sec.~\ref{nononjet} we demonstrated that whereas the nonjet quadrupole is a necessary and unique element of the standard fit model independent of the SS 2D peak a nonjet sextupole is not a necessary element. A NJ sextupole component cannot be isolated (with systematic significance) from the SS 2D peak structure. What is uniquely defined is the SS 2D peak relative to fitted AS dipole and NJ quadrupole within the standard fit model. The structure of the SS peak is factorizable, and each factor can be represented by a functional form. That unique structure can then be compared to theoretical predictions in whole or in parts according to explicit hypothesis.

\subsection{Modeling data with extreme conditions imposed} \label{ultra}

For almost all 2D angular correlations from RHIC collisions the data histograms within some limited $\eta$ acceptance can be described by the ``standard fit model'' consisting of an away-side 1D dipole, a nonjet quadrupole and a same-side 2D peak. We have shown that any ``higher harmonics'' must then originate with the SS 2D peak. But for a small fraction of the data, associated with what may be called ``extreme conditions,'' angular correlation structure may deviate substantially from the standard fit model. Such conditions include high-$p_t$ ``trigger-associated'' cuts, cuts favoring large \pp collision multiplicities, acceptance cuts emphasizing large $\eta$ values and ``ultra-central'' \aa collisions (0-1\% central).

For \pp and more-peripheral \auau collisions the entire SS 2D peak is resolved within the STAR TPC angular acceptance and the correlation model is fully constrained. The SS 2D peak model is unambiguous: a 2D Gaussian. The SS peak continues to be well resolved until well above the sharp transition in peak properties at $\nu_{trans}$~\cite{anomalous}. However, in more-central collisions the elongated SS 2D peak extends sufficiently far outside the TPC $\eta$ acceptance that its modeling on $\eta_\Delta$ becomes ambiguous.  When only a fraction of the SS peak is observed some peak properties (e.g., amplitude and curvature at the mode) are more reliably determined than others (e.g., higher moments, tail structure). If a greater fraction of the SS 2D peak becomes accessible within a larger detector  acceptance more peak properties can be determined, including possible non-Gaussian tails at larger $\eta_\Delta$. 

Substantial deviations of same-side and away-side peak structure from the standard fit model are observed at the RHIC and LHC for certain combinations of the extreme conditions listed above. The SS 2D peak acquires long tails on $\eta_\Delta$, and the AS 1D peak develops a substantial deformation.  All such deviations can be described in terms of changes in the SS 2D or AS 1D peak model functions. In the former case the change appears restricted to the 1D $H(\eta_\Delta)$ factor, and modified Gaussian or non-Gaussian peaked functions are available (e.g., Sec.~IV-C of Ref.~\cite{tzyam}). 

But a trend has emerged to decompose the SS 2D peak into ``jet-like'' (small $\eta_\Delta$) and ``ridge-like'' (large $\eta_\Delta$) components and to model the latter with a 1D Fourier series {\em in common with all other correlation structure} (i.e., nonjet quadrupole and AS 1D peak). Single Fourier terms may actually represent multiple physical mechanisms, including jet formation. However, the convention is adopted to interpret each term as a harmonic flow, including the $m=1$ dipole. 
In that system the connection with established jet structure in peripheral collisions is broken, and there can be no consistent description of centrality evolution. Consistency could be restored by acknowledging that whatever the imposed conditions, modeling the SS 2D and AS 1D peaks  separately is essential. 
A single 1D Fourier series representing two peaked structures plus other nonjet components is not interpretable.

\section{Summary} \label{sum}

The information contained in 2D angular correlation data from nuclear collisions, both $p_t$-integral and $p_t$-differential, is extensive and could be used to eliminate invalid theoretical models and suggest valid physical interpretations.  However, some analysis methods minimize and distort that information by (a) projection onto subspaces and (b) many-to-one and one-to-many mappings between correlation components and model elements.

Recently, a major initiative has emerged to infer ``higher harmonic flows'' (azimuth multipoles) from correlation data. Fourier analysis of angular correlations projected onto 1D azimuth from limited pseudorapidity intervals is used to infer such multipoles. However, 1D Fourier analysis conflicts with 2D model fits to unprojected angular correlations in which apparent jet-like structure appears to dominate correlations. To help resolve the impasse we have conducted a detailed study of 2D angular correlation analysis methods and interpretations relating to the question of ``higher harmonic flows.''

A 1D Fourier analysis of 2D angular correlations projected onto azimuth must describe the periodic data accurately by definition. But there is nothing to compel interpretation of the series terms as flows. The unprojected 2D structure can be decomposed precisely into several components that retain their identity from \pp collisions to central \auau collisions. One of those components, the same-side 2D peak,  is strongly structured on pseudorapidity. A single 1D Fourier series cannot model that component and therefore {\em fails as a data model}. More-detailed examination establishes that multiple correlation components contribute to some individual Fourier terms (many-to-one) and some correlation components contribute to multiple Fourier terms (one-to-many) leading to severe interpretation difficulties. Further analysis establishes that ``higher harmonics'' inferred from 1D Fourier analysis are at least dominated by the SS 2D (jet-related) peak.

A potential remedy for such ambiguities might be achieved by adding higher multipole terms to the ``standard fit model'' of angular correlations that successfully describes \pp and more-peripheral \auau correlations. However, we demonstrate in this study that additional multipole elements can conspire to contribute to the jet-related SS 2D peak model rather than representing sought-after {\em nonjet} multipoles that might be interpreted as flows. For more-central \auau data adding $m > 2$ terms to the standard data model leads to {\em systematically insignificant} changes in the SS 2D peak model. Although some model parameters may change substantially the overall data model (sum of the elements) is quite stable and provides a unique 2D data description.

The last part of the study was determination of the {\em necessity} for nonjet higher multipoles in the data description. A known sextupole contribution was added to a data histogram and fitted with the standard 2D data model. The result was a change in the inferred SS 2D peak structure but no appearance of the sextupole component in the fit residuals. In contrast, omission of the nonjet quadrupole element from the fit model produced an equivalent component in the fit residuals. The nonjet quadrupole is thus a necessary and distinct model element. But we conclude that it is impossible to isolate higher nonjet multipoles from the SS 2D peak structure.

It has been argued that at least in more-central \aa collisions some part or all of the SS 2D peak is a flow phenomenon. Its Fourier components might then be interpreted legitimately as flow manifestations. However, a general model-independent analysis program should confront the {\em intact} structure of the SS 2D peak to determine the correct $H(\eta_\Delta)$ functional form vs systematic conditions. Theory tests must accommodate the SS 2D peak in its entirety, with all its related phenomenology, or fail. 

This work was supported in part by the Office of Science of the U.S. DOE under grants DE-FG03-97ER41020 (UW) and DE-FG02-94ER40845 (UTA).


\end{document}